\magnification=1200

\hsize 17truecm
\vsize 23truecm

\font\twelvec=msbm10 at 10pt
\font\sevenc=msbm10 at 7pt
\font\fivec=msbm10 at 5pt

\newfam\co
\textfont\co=\twelvec
\scriptfont\co=\sevenc
\scriptscriptfont\co=\fivec

\def\arg{\mathop{\rm arg}\nolimits}
\def\Const{\mathop{\rm Const.}\nolimits}
\def\det{\mathop{\rm det}\nolimits}
\def\exp{\mathop{\rm exp}\nolimits}

\def\Id{\mathop{\rm Id}\nolimits}

\def\lim{\mathop{\rm lim}\nolimits}

\def\mod{\mathop{\rm mod}\nolimits}

\def\sgn{\mathop{\rm sgn}\nolimits}

\def\neigh{\mathop{\rm neigh}\nolimits}

\def\Sum{\displaystyle\sum}
\def\e{\mathop{\rm \varepsilon}\nolimits}

\baselineskip 15pt

\centerline {\bf THE SEMI CLASSICAL MAUPERTUIS-JACOBI CORRESPONDANCE} 

\centerline {\bf FOR QUASI-PERIODIC HAMILTONIAN FLOWS}
\bigskip
\centerline{Sergey DOBROKHOTOV ${}^{(*)}$ {\it\&} Michel ROULEUX ${}^{(**)}$}
\bigskip
\centerline {${}^{(*)}$ Institute for Problems in Mechanics of Russian Academy of Sciences}

\centerline {Prosp. Vernadskogo 101-1, Moscow, 119526, Russia, dobr@ipmnet.ru} 

\centerline {${}^{(**)}$ Centre de Physique Th\'eorique and Universit\'e du Sud Toulon-Var, UMR 6207}

\centerline {Campus de Luminy, Case 907, 13288 Marseille Cedex 9, France, rouleux@cpt.univ-mrs.fr}
\bigskip
\noindent {\bf Abstract}: We extend to the semi-classical setting the Maupertuis-Jacobi correspondance for a pair of 
hamiltonians $(H(x,hD_x), {\cal H}(x,hD_x)$. If ${\cal H}(p,x)$ is completely integrable, or has merely an invariant diohantine torus 
$\Lambda$ in energy surface ${\cal E}$, then we can construct a family of quasi-modes for $H(x,hD_x)$ at the corresponding 
energy $E$. This applies in particular to the theory of water-waves in shallow water, and determines trapped modes
by an island, from the knowledge of Liouville metrics.
\medskip
\noindent{\bf 1. Introduction}.
\medskip
Let $M$ be a $d$ dimensional Riemannian manifold, with metric tensor 
$ds^2=\Sum_{1\leq i,j\leq d}g_{ij}(x)$ $dx_i\otimes dx_j$
and consider a particle of unit mass moving along geodesics of $M$ with kinetic energy $T={1\over 2}\bigl({ds\over d\tau}\bigr)^2$.
Suppose we are now given a potential $V(x)$. There is a relation between the geodesic flow 
and the Hamiltonian flow parametrized by $t\in{\bf R}$,
for Hamiltonian $H(p,x)={1\over 2}\Sum_{1\leq i,j\leq d}g^{ij}(x)p_ip_j+V(x)$ on the energy surface $E$. 
To find the dependence $(P(t),X(t))$ on the new time $t$ one has to solve the equation
$dt=(2(E-V))^{-1/2}ds$
which we call Maupertuis principle (see [Ar2]). Then the particle moves along the geodesic flow on $M$ relative to 
the metric 
$d\rho=(E-V)^{1/2}ds$, and we say that both flows (and metrics) are {\it conformally equivalent}.

We shall consider in this paper the following stronger variant.
Maupertuis Jacobi correspondance for the pair  
$(H, {\cal H})$, ${\cal H}(p,x)={1\over 2(E-V)}\Sum_{1\leq i,j\leq d}g^{ij}(x)p_ip_j$, 
ensuring that both systems have the {\it same} trajectories $(P(t),X(t))$ and $({\cal X}(\tau), {\cal P}(\tau))$
at energies $(E,1)$, is given by $d\tau=(E-V)dt$.
This principle was used for instance by T.Levi-Cevita in relation with Kepler problem, cf. [Ts], [HieGrDoRa] and references therein~;
see also [EasMat] for the case of {\it projective equivalence} of metric connexions. 
Among interesting situations of application of Maupertuis-Jacobi correspondance is the case,  
when the functions $({\cal X}(\tau), {\cal P}(\tau))$ are 
$\tau$-quasiperiodic~: $({\cal P}(\tau),{\cal X}(\tau))=({\cal P}(\omega\tau+\varphi^0)),{\cal X}(\omega\tau+\varphi^0))$.
Here $\omega$ is a vector of periods and $({\cal P}(\tau),{\cal X}(\tau))$ are $2\pi$-periodic functions with respect to each phase. 

This occurs when ${\cal H}$ is completely integrable, and $\Lambda$ are compact, connected leaves 
of maximal dimension $d$ (i.e. lagrangian) in the fibration of $T^*M$ by the momentum map $m$, that we 
assume to be non singular at $\Lambda=\Lambda_0$. Then Arnold-Mineur-Liouville theorem states [Ar2] that in a suitable 
action-angle coordinate system, $\Lambda$ are diffeomorphic to the torus ${\bf T}^d={\bf R}^d/2\pi{\bf Z}^d$, and parametrized   
by the action variables $I\in{\bf R}^d_+$. The non-criticality of $m$ implies that all components $I_j$ of $I$ are
non zero in a neighborhood of $\Lambda_0$. 

In general the new hamiltonian $H$ has no reason to be integrable. Nevertheless, 
in a neighborhood of $\Lambda_0$, there are action-angle coordinates $(J,\varphi)$ giving a full family of tori $\Lambda^J$
which are almost invariant for the hamiltonian vector field $X_H$. Under some diophantine condition, 
it it possible to improve these tori, so that they become invariant within an arbitrary accuracy.

This is actually the case, for using a generalization of Birkhoff transformations (better known near a stationary   
point of $X_H$,~) in the symplectic coordinates given by Darboux-Weinstein theorem, $H$ can be brought to a normal form,
we call the Birkhoff normal form (BNF). This was carried out by M.Hitrik, J.Sj\"ostrand and S.Vu-Ngoc [HiSjVu].

An alternative, and more pragmatic way to our opinion, was devised in [BeDoMa], and consists in providing directly suitable action-angle 
coordinates $(J,\varphi)$, without resorting to Darboux-Weinstein theorem. We believe it has the advantage of being easily
implemented numerically in some concrete applications.

Once the classical Hamiltonian is taken to BNF, we may address the problem of finding quasi-modes supported on the $\Lambda^J$,
with a suitable accuracy. A necessary condition for $\Lambda^J$ is to satisfy Maslov quantization condition. 
A similar problem arises in an attempt of quantizing the so-called KAM tori. Quoting J.J. Duistermaat [Du1,p.231],
while slightly changing notations~: 

``\dots there is no reason why the tori on which [Maslov quantization condition] holds for some $h>0$ should persist. For most 
[invariant embedded lagrangian manifolds $\iota:\Lambda\to T^*M$] one can expect that $h\mapsto {1\over  h}\cdot \theta$
[the cohomology class of the closed 1-form $\iota^*(pdx)$] is dense in $H^1(\Lambda;{\bf R})/H^1(\Lambda;{\bf Z})$. 
Extracting a subsequence ${1\over  h}=\tau_{n_k}$, from $\tau_n=\tau_0+n\sigma_0, n\in{\bf Z}$ 
[which ensures Maslov quantization condition to hold], such that $\tau_n\cdot\theta$ converges to $-\alpha/4$
[Maslov class] in $H^1(\Lambda;{\bf R})/H^1(\Lambda;{\bf Z})$, one can still obtain some asymptotic estimates for the
spectrum of $H$ but in general the convergence of $\tau_n\cdot\theta$ to $-\alpha/4$ is expected to be slow 
and the asymptotic estimates for the spectrum are correspondingly weak.''

This observation was made more precise by Y.Colin de Verdi\`ere [CdV,Prop.7.5]~: namely, if $H^1(\Lambda;{\bf R})$ is of dimension 2
(i.e. for a 2-d torus $\Lambda$), then the extraction above can be made, provided $\theta$, expressed in a basis of entire classes,
is irrational.

A different point of vue consists in selecting among the family $\Lambda^J$ suitable tori
on which Maslov quantization condition holds, i.e. a sequence $J=J_k(h), k\in{\bf Z}, |k|h\leq h^\delta$.
This allows to produce an almost ``full'' (or ``density 1'') spectral sequence for $H$, in some suitable sense.

The first route, followed in [HiSjVu], consists, through $h$-Pseudo Differential Calculus, in quantizing the BNF microlocally near the  
$\Lambda^J$, and this can be achieved for corrections to all orders in $h$. 
(See also [CdV1], [Laz], [Po], and [Vu] for the case of a family of $d$ commuting $h$-PDO.~)

This is a rather remarkable result, since quantization of manifolds is in general impossible beyond first order. 
It relies essentially on the fact, observed a long time ago by Colin de Verdi\`ere [CdV], 
and used later by G.Popov [Po], that the action of $H$ on semi-densities on
tori verifying Maslov quantization condition, reduces to the action of another $h$-PDO, obtained by conjugating $H$
by a Fourier Integral Operator (FIO) quantizing the BNF, to microlocally defined functions on the flat torus
${\bf T}^d$. These functions satisfy Floquet periodicity condition~:
$$u(x-\nu)=e^{2i\pi\langle I+\alpha h/4,\nu\rangle/h}u(x),  \quad \nu\in {\bf Z}^2\leqno(0.3)$$
Here $I=(I_1,\cdots,I_d)$ are the classical actions parametrizing the torus $\Lambda$
(stated otherwise, $I$ is a representant of Liouville class $\theta$  in $H^1(\Lambda;{\bf R})$), so that 
$u(x-\nu)=u(x)$ for all integer $\nu$ whenever $\theta/h+\alpha/4=0$ mod ${\bf Z}^2$. 

Here we have chosen another route, consisting simply in using the almost invariant tori $\Lambda^J$ to
construct quasi-modes for operator
$H(x,hD_x)$ near energy 0. This can be done by adjusting the new action variable $J$ so to 
satisfy Maslov quantization condition, and solving the transport equations for $H$ to first order along $\Lambda^J$.
We believe again that these constructions are simpler from a pragmatic
point of view, especially when $H$ assumes
the form $H=-h^2\Delta+V$. In more complex situations however where  (see Section 4.d) we have to resort to FIO's
implementing the change of variables that takes our operator to the normal form, passing for instance from a general
conformal metric to a model one, but the constructions are very similar.

We could also include isotropic tori, 
corresponding to the gaps in the foliation of the energy surface by lagrangean manifolds when the
system is no longer integrable, or consider critical points of the moment map, which
leads to introduce so-called harmonic oscillator coordinates (in the elliptic case).

The paper goes as follows. In Sect.1 we recall from [BeDoMa] the construction of action-angle coordinates in the 
neighborhood of a torus $\Lambda$ invariant for the hamiltonian flow $X_H$, leading to existence of a family of tori 
$\Lambda^{J',N}$ in the neighborhood of $\Lambda$ which, if $\Lambda$ is diophantine,  are quasi-invariant over arbitrarily long times.
Thus we get a variant of BNF. 

In Sect.2 we quantize some of these $\Lambda^{J',N}$, so that they satisfy Bohr-Sommerfeld-Maslov quantization condition, and
support quasi-modes for $H$ with a suitable accuracy.

In Sect.3 we apply these considerations to semi-classical Jacobi-Maupertuis correspondance~: 
assuming that ${\cal H}$ is completely integrable,
or simply has an invariant torus $\Lambda$ in some energy surface, we show how to construct quasi-modes for 
$H$. We show that the frequencies $\widetilde\omega$ (resp. $\omega$) 
of the quasi-invariant flows for ${\cal H}$ (resp. $H$) are simply related by $\omega=\langle g\rangle\widetilde\omega$.
This implies partial isospectrality of both hamiltonians.

In Sect.4 we give various applications of construction of action-angle variables, and examples for
semi-classical Jacobi-Maupertuis correspondance, including the shallow-water waves theory. 
Namely we show how to select profiles of the basin, corresponding to a family of Liouville 2-D integrable metrics,
which give raise to trapped modes of a given wavelength. This could be easily implemented on a computer program, and 
consists actually in the main ``practical'' result of that paper.

In Appendix, we give a short account for Maslov theory and WKB constructions as needed in the main text. 

\medskip
\noindent{\bf Acknowledgments}: One of us (M.R.) thanks Ch.Duval, J.Sj\"ostrand and D.H\"afner for useful remarks.
We also acknowledge a grant from the R\'egion Provence-Alpes-C\^ote d'Azur in 2007, which enabled 
to spend some fruitful and enjoyable time in both institutions.

\bigskip
\noindent {\bf 1. Action-angle variables for a family of almost invariant tori.}
\medskip
First we recall the main result of [BeDoMa]. 
Consider a smooth Lagrangian torus 
$\Lambda_0=\{(p,x)=(P^0(\varphi),X^0(\varphi)),\ \varphi=(\varphi_1,\cdots,\varphi_d)\in{\bf T}^d\}$,
${\bf T}^d=\bigl({\bf R}/2\pi{\bf Z}\bigr)^d$
embedded in $T^*{\bf R}^d$, and let $I^0=(I^0_1,\cdots,I^0_d)$,
$$I^0_j={1\over  2\pi}\oint_{\gamma^0_j} P^0 d X^0\leqno(1.1)$$
be the corresponding actions computed along the basis of cycles
$(\gamma^0_1,\cdots,\gamma^0_d)$ on $\Lambda_0$. We denote by $Y^0(\varphi) $ the $2d$ vector function ${}^t(P^0,X^0)$. 
Although it is possible to take degeneracies into account,
we shall assume here that the moment map is non singular near $\Lambda_0$, i.e. $I^0_j\neq 0$ for all $j$. 

We may also assume that $\Lambda_0$ belongs to a $\ell$-parameter family of tori $\Lambda^s$, $s$ varying in an open set of 
${\bf R}^\ell$, $1\leq\ell\leq d$. This is the case, by Arnold-Liouville-Mineur theorem, if these tori arise from an integrable,
or semi-integrable Hamiltonian system. If $\ell=1$ we can think of $s=E$ as the energy of the system, 
but more generally,
the family can be parametrized (possibly after renumbering variables,~) by
the first $\ell$ action variables $(I_1,\cdots,I_\ell)$. 
However, tori of the family $\Lambda^s$ are not necessarily preserved by the transformations below, so to avoid technical difficulties
we shall here take simply $\ell=0$, and consider a single torus $\Lambda_0$. 

Introduce the 
$2d\times d$ matrix $Y^0_{\varphi}$ with columns $(Y^0_{\varphi_1},\cdots,Y^0_{\varphi_d})$,
and the $d\times d$ symmetric matrix
$M=-({}^t Y^0_\varphi Y^0_\varphi)^{-1}$ (we denote often partial derivatives by a subscript.~)
Let also ${\cal J}$ be the standard symplectic matrix ${\cal J}=
\normalbaselineskip=18pt\pmatrix{0 & -{\Id }_d \cr {\Id }_d & 0 \cr}$, where  
${\Id }_d$ is the $d\times d$ unit matrix. We put 
$$Y^J={P^J(\varphi)\choose X^J(\varphi)}={\cal J} Y^0_\varphi M \leqno(1.2)$$
The $2d\times d$ matrices $Y^0_{\varphi}$ and $Y^J$ satisfy the conditions
$${}^tY^0_\varphi{\cal J}Y^0_\varphi=0, \quad
{}^tY^J {\cal J}Y^0_\varphi =-{\Id }_d, \quad  {}^tY^J{\cal J}Y^J=0,\leqno(1.4)$$
the last two being implied by (1.2) and the first one, which precisely express that $\Lambda_0$ is a Lagrangian torus.
This is proved in Appendix of [BeDoMa] by a direct computation. 

In the notation $Y^J$, index $J$ refers to new action variables we are constructing below,
for which $Y^J$ enjoys the same symplectic properties as ${\partial Y\over\partial I_j}$ for an integrable system. 
All these matrices are periodic with respect to $\varphi\in{\bf T}^d$. 

Now we introduce precisely the new action-angle coordinates $(J,\varphi)\in\neigh (I^0,{\bf R}^d)\times\bigl({\bf R}/2\pi{\bf Z}\bigr)^d$,
so that $y'=Y(J,\varphi)={P(J,\varphi)\choose X(J,\varphi)}$ 
are canonical coordinates defined in a neighborhood of $\Lambda_0$.
As is shown in [BeDoMa] this can be achieved by taking 
$$Y(J,\varphi)=Y^0(\varphi)+Y^J(\varphi)G(J,\varphi)\leqno(1.5)$$
where the real $d$-vector $G={}^t(G_1,\cdots,G_d)$ is obtained by solving the system of the quadratic equations~:
$$G_j-{1\over 2}\langle G,{}^tY^J{\cal J}Y^J_{\varphi_j}G\rangle=\iota_j, \; j=1,2,\dots,d\leqno(1.6)_j$$
with data $\iota=(\iota_1,\cdots,\iota_d)=J-I^0$. We fix the root of this equation by the requirement
$G(J,\varphi)=\iota+{\cal O}(\iota^2)$ as $\iota\to0$. Equations (1.6) can be solved with the help of a fixed point theorem, and  
$G$ depends smoothly and periodically in  $\varphi$, analytically in $J$ (or $\iota$).  For completeness we recall  the following~:
\medskip
\noindent{\bf Lemma 1.1} [BeDoMa]:
Coordinates $y'=(p',x')=y+{\cal O}(\iota)$ are symplectic, i.e. they verify
$$\langle  y'_{\varphi_j},{\cal J} y'_{\varphi_k}\rangle=\langle  y'_{J_j},{\cal J} y'_{J_k}\rangle=0, \quad
\langle  y'_{J_j},{\cal J} y'_{\varphi_k}\rangle=-\delta_{jk}, \quad j=1,\dots,d \leqno(1.7)$$
\smallskip
\noindent{\it Proof}: Consider first $\langle  y'_{J_j},{\cal J} y'_{J_k}\rangle$.   
Differentiating (1.5) w.r.t. $J$ and making use of the last relation (1.4) we find 
$\langle  y'_{J_j},{\cal J} y'_{J_k}\rangle=\langle G_{J_j},{}^tY^J{\cal J}Y^JG_{J_k}\rangle=0$.

Consider next $A_{jk}=\langle  y'_{J_j},{\cal J} y'_{\varphi_k}\rangle$.
Differentiating (1.5) w.r.t. $\varphi_k$ and $J_j$, then making use of the second relation (1.4) in the form 
${}^tY^J {\cal J}Y^0_{\varphi_k}=-e_k$, $e_k$ being the $k$:th unit vector of ${\bf R}^d$, and of the last relation (1.4), we find
$$A_{jk}=-G_{kJ_j}+\langle G_{J_j},{}^tY^J{\cal J}Y^J_{\varphi_k}G\rangle$$
Then differentiating (1.6$)_j$ w.r.t. $J_k$, and making use of the relation 
${}^tY^J{\cal J}Y^J_{\varphi_j}=-{}^tY^J_{\varphi_j}{\cal J}Y^J$ which follows from the last equation (1.4), 
gives $A_{jk}=-\delta_{jk}$ as stated. 

At last, consider $B_{kj}=\langle  y'_{\varphi_k},{\cal J} y'_{\varphi_j}\rangle$, which by 
(1.5) again gives 
$$B_{kj}=\langle{\cal J}Y^0_{\varphi_j}+{\cal J}Y^J_{\varphi_j}G+{\cal J}Y^JG_{\varphi_j},
Y^0_{\varphi_k}+Y^J_{\varphi_k}G+Y^JG_{\varphi_k}\rangle\leqno(1.8)$$
Differentiating $(1.6)_j$ w.r.t. $\varphi_k$, then $(1.6)_k$ w.r.t. to $\varphi_j$, 
and forming the difference, the mixed derivatives $Y^J_{\varphi_k\varphi_j}$ cancel out, which gives~:
$$\eqalign{
-G_{j\varphi_k}&+G_{k\varphi_j}+{1\over 2}\langle G_{\varphi_k},{}^tY^J{\cal J}Y^J_{\varphi_j}G\rangle
-{1\over 2}\langle G_{\varphi_j},{}^tY^J{\cal J}Y^J_{\varphi_k}G\rangle+\cr
&{1\over 2}\langle G,\bigl({}^tY^J_{\varphi_k}{\cal J}Y^J_{\varphi_j}-{}^tY^J_{\varphi_j}{\cal J}Y^J_{\varphi_k}\bigr)G\rangle
+{1\over 2}\langle G,{}^tY^J{\cal J}\bigl(Y^J_{\varphi_j}G_{\varphi_k}-Y^J_{\varphi_k}G_{\varphi_j}\bigr)\rangle=0\cr
}$$
We have ${}^tY^J_{\varphi_k}{\cal J}Y^J_{\varphi_j}-{}^tY^J_{\varphi_j}{\cal J}Y^J_{\varphi_k}=
2{}^tY^J_{\varphi_k}{\cal J}Y^J_{\varphi_j}$, and 
$\langle G,{}^tY^J{\cal J}Y^J_{\varphi_j}G_{\varphi_k}\rangle=\langle G_{\varphi_k},{}^tY^J{\cal J}Y^J_{\varphi_j}G\rangle$,
so
$$-G_{j\varphi_k}+G_{k\varphi_j}+\langle G_{\varphi_k},{}^tY^J{\cal J}Y^J_{\varphi_j}G\rangle
-\langle G_{\varphi_j},{}^tY^J{\cal J}Y^J_{\varphi_k}G\rangle+
\langle G,{}^tY^J_{\varphi_k}{\cal J}Y^J_{\varphi_j}G\rangle=0$$
Replacing the term $\langle G,{}^tY^J_{\varphi_k}{\cal J}Y^J_{\varphi_j}G\rangle$ in (1.8) by its expression, 
we see that $\langle G_{\varphi_j},{}^tY^J{\cal J}Y^J_{\varphi_k}G\rangle$
cancels out with $\langle {}^tY^J_{\varphi_k}{\cal J}Y^JG_{\varphi_j},G\rangle$,
$\langle G_{\varphi_k},{}^tY^J{\cal J}Y^J_{\varphi_j}G\rangle$ cancels out with
$\langle {}^tY^J{\cal J}Y^J_{\varphi_j}G,G_{\varphi_k}\rangle$,
while $\langle {}^tY^J{\cal J}Y^JG_{\varphi_j},G_{\varphi_k}\rangle=\langle {\cal J}Y^0_{\varphi_j},Y^0_{\varphi_k}\rangle=0$
by (1.4). So (1.8) simplifies into
$$B_{kj}=G_{j\varphi_k}-G_{k\varphi_j}+\langle{\cal J}Y^J_{\varphi_j}G,Y^0_{\varphi_k}\rangle
+\langle{\cal J}Y^JG_{\varphi_j},Y^0_{\varphi_k}\rangle+\langle{}^tY^J_{\varphi_k}{\cal J}Y^0_{\varphi_j},G\rangle
+\langle{}^tY^J{\cal J}Y^0_{\varphi_j},G_{\varphi_k}\rangle$$
and because again of ${}^tY^J{\cal J}Y^0_{\varphi_k}=-e_k$, $B_{kj}=0$. The Lemma is proved. $\clubsuit$. 
\smallskip
In other words, the transformation $\kappa :(J,\varphi)\mapsto y'=\kappa(J,\varphi)$ is canonical.
This yields a $d$-parameter family of tori $\Lambda^J$, by choosing $J=I^0+\iota$= Const., and 
$\Lambda^J=\Lambda_0+{\cal O}(\iota)$ in the $C^\infty$ topology. 

\noindent {\it Remark}: Actually we know, by Darboux-Weinstein theorem (see [BaWe,Corollary  4.20])
that if $\Lambda\subset T^*{\bf R}^d$ is a lagrangian submanifold, then a tubular neighborhood of $\Lambda$ in $T^*{\bf R}^d$
is symplectomorphic to a neighborhood of the zero section $Z\approx\Lambda$ in $T^*\Lambda$, by a map which is the identity on
$\Lambda$. The construction above specializes and provides a simple proof to Darboux-Weinstein theorem in the case where $\Lambda$
is a torus given by $J=I^0$ in action-angle coordinates, so that neighborhoods of $Z$ in $T^*\Lambda$ can be identified
with $\neigh (I^0;{\bf R}_+^d)\times {\bf T}^d$. In other words, given the lagrangian immersion $\iota:\Lambda\to T^*M$,
we have constructed a symplectic immersion ${\cal I}$ of a neighborhood of $Z$ into $T^*M$,
such that ${\cal I}=\iota\circ\pi_\Lambda$ on $Z$, where $\pi_\Lambda:T^*\Lambda\to\Lambda$ is the natural projection.
\medskip
We still denote by $\gamma_j$ the basic cycles on $\Lambda^J$,
which are obtained by a continuous deformation of those on $\Lambda$, because of the nondegeneracy condition.
So we have
$$J_j={1\over 2\pi}\oint_{\gamma_j} p'dx'\leqno(1.10)$$
Assume now that $\Lambda_0$ is an integral manifold for Hamiltonian $H(p,x)$, i.e. 
$p=P^0(\varphi+\omega t), x=X^0(\varphi+\omega t)$ is a quasi-periodic solution of 
the hamiltonian system $(\dot p,\dot x)=X_H(p,x)$ on $\Lambda_0$ with rationally independent frequencies $\omega$,
we try to construct a normal form for $H$ near $\Lambda_0$. Here notation $X_H(p,x)$ stands for the Hamiltonian vector field of $H(p,x)$.
By a Taylor expansion in coordinates $y'$ as above, we have~:
$$H\circ\kappa(J,\varphi)=H\bigl(Y^0(\varphi)+Y^J(\varphi)G(J,\varphi)\bigr)=H(Y^0(\varphi))+\langle H'(Y^0(\varphi)),
Y^J(\varphi)G(J,\varphi)\rangle +{\cal O}(\iota^2)\leqno(1.11)$$
with $H'(Y^0(\varphi))=-{\cal J}\dot y=-{\cal J}\langle \omega,{\partial\over\partial\varphi}\rangle Y^0(\varphi)$. So 
$$\langle H'(Y^0(\varphi)),Y^J(\varphi)G(J,\varphi)\rangle=\Sum_{j,m}\omega_jY^0_j(\varphi)Y_m^J(\varphi)G_m(J,\varphi)$$
which by (1.4) and the relation $G(J,\varphi)=\iota+{\cal O}(\iota^2)$ gives 
$$H(y')=H_0+\langle\omega,\iota\rangle+{\cal O}(\iota^2),\qquad H_0= 
H\big|_{\Lambda_0}\leqno(1.12)$$
so the $\Lambda^J$ are invariant at first order in $\iota$ under the flow of $H\circ\kappa$.

Assuming further a diophantine condition on $\omega$, we can extend this normal form as follows, 
using a finite order KAM procedure. We look for a generating function $S(J', \varphi)$ such that
$J=\partial_\varphi  S(J',\varphi)$, $\varphi'=\partial_{J'}S(J',\varphi)$,  where $J'=I^0+\iota'$.
We find that $S(J',\varphi)=\langle J',\varphi\rangle+\Phi(\iota',\varphi)$ is defined as a Fourier  series
in  $\varphi\in{\bf T}^d$, with vanishing zero mode, and coefficients $\Phi_k(\iota')={\cal O}(\iota'^2)$.
Push Taylor expansion (1.11) to next order in $\iota=\iota'+\partial_\varphi\Phi(\iota',\varphi)$,
we find after averaging over $\varphi$ the quadratic terms in $\iota'$ 
$$H(P(J,\varphi),X(J,\varphi))=H_0+\langle\omega,\iota\rangle+A(\iota')+B(\iota',\varphi)+
\langle \omega,\partial_\varphi\Phi(\iota',\varphi)\rangle+
{\cal O}(\iota'^3)\leqno(1.13)$$
where $A(\iota'), B(\iota',\varphi)={\cal O}(\iota'^2)$ are given smooth functions  
defined in a neighborhood of  $\Lambda_0$,
and $\langle  B(\iota',\cdot)\rangle=0$ (average over ${\bf T}^d$.~)
Since $\omega$ is diophantine, and $B(\iota',\varphi)$ smooth in $\varphi$, 
we can determine the quadratic part of $\Phi$ in $\iota'$ by
solving $B(\iota',\varphi)+\langle \omega,\partial_\varphi\Phi(\iota',\varphi)\rangle=0$ up to  ${\cal O}(\iota'^3)$ terms. 
This process easily carries out at any given order, yielding the following~:
\medskip
\noindent {\bf Theorem 1.2} [BeDoMa]: Let $H$ be as above, and assume the frequency vector $\omega$ satisfies a diophantine
condition. Then for any $N=1,2,\cdots$, there is a smooth canonical map $\kappa_N:(J',\varphi')\mapsto (J,\varphi)$ defined for  
$J'$ in  a neighborhood of $I$  and $\varphi'\in{\bf T}^d$, and a polynomial $H_N(\iota')$ of degree $N$,
with $H_N(\iota')=H_0+\langle\omega,\iota'\rangle+{\cal O}(\iota'^2)$, $H_0=\langle \omega,I\rangle$ such that 
$$H\circ\kappa\circ\kappa_N(J',\varphi')=H_N(\iota')+{\cal O}(\iota'^{N+1})\leqno(1.15)$$
\smallskip
This is precisely a Birkhoff normal form for $H$ in a neighborhood of $\Lambda_0$ (see also [HiSjVu].~)
Assuming analyticity on $H$, these estimates could certainly be sharpened to give a remainder 
${\cal O}(e^{-\iota'^{-\beta}/C})$ as in Nekhoroshev theorem, $0<\beta<1$ being related to the diophantine condition,
and $C>0$ (see e.g. [Po2].~)
\smallskip
To conclude this Section,  we make a few remarks concerning quantization in action-angle variables, following [BeDoMa]. 
Recall [BaWe, Definition 5.33] that a linear map $\rho$
from the space $C^\infty(T^*M)$ of smooth functions, to the algebra ${\cal A}$ 
generated by self-adjoint operators on some complex Hilbert space ${\cal H}$, and endowed with the Lie algebra structure defined by
$[A,B]_h={i\over h}(AB-BA)$, is called a {\it quantization} 
provided it satisfies so-called Dirac axioms~: (1) $\rho(1)=\Id$, (2) $\rho(\{f,g\})=[\rho(f),\rho(g)]_h$, (3) for some complete set of 
functions $f_1,\cdots,f_n$ in involution, the operators $\rho(f_1),\cdots,\rho(f_n)$ form a complete commuting set.
We know this set of axioms is in general too stringent, in the sense that a quantization of all classical observables doesn't exist,
although the symbolic calculus of $h$-Fourier integral operators, or Maslov canonical operators, provides a fairly good approximation
of this classical-quantum correspondence. 

Quantization deformation occurs already in the simple case of
a completely integrable hamiltonian system on $T^*M$, with hamiltonian $H(p,x)$, 
which admits a family of lagrangian tori $\Lambda^I$. Namely, trying to quantize the corresponding 
action-angle variables $(I,\varphi)$, considered as classical observables, in a neighborhood of the $\Lambda^I$'s, we
require that $\rho(I)=\widehat I,\rho(\varphi)=\widehat\varphi$ would satisfy 
$$[\widehat I_j,\widehat I_k]=0, \quad [\widehat \varphi_j,\widehat I_k]_h=\delta_{jk}, 
\quad [\widehat \varphi_j,\widehat \varphi_k]_h=0\leqno(1.17)$$
and moreover, that the semi-classical hamiltonian $\widehat H(x,hD_x)$ associated with $H(x,p)$ {\it via} usual 
Weyl $h$-quantization, would be a function of $(\widehat\varphi,\widehat I)$. The naive answer would consist in choosing 
$\widehat\varphi$ as multiplication by $\varphi$, and $\widehat I=hD_\varphi$. 
But then if we try to recover the canonical operators 
$\widehat x_j=X_j(\widehat\varphi,\widehat I),\widehat p_k=P_k(\widehat\varphi,\widehat I)$ by symbolic calculus,
it turns out that the canonical commutation relations 
$[\widehat p_k,\widehat x_j]_h=\delta_{jk}$ are only satisfied modulo ${\cal O}(h)$.  (This can be easily checked also when $H$
is the harmonic oscillator, and $x=\sqrt I\cos\varphi,x=\sqrt I\sin\varphi, H(p,x)=p^2+x^2=I$.~) Therefore 
one should in general also include quantum corrections and seek for canonical variables as 
$\widehat x_j=X_j(\widehat\varphi,\widehat I)+hX_j^{(1)}(\widehat\varphi,\widehat I)+\cdots$. This in turn affects the full
symbol of hamiltonian $H$, expressed in action-angle variables, by a quantity ${\cal O}(h^2)$, and
$H(p,x,h)=H_0(I)+h^2H_1(I,\varphi)+\cdots$ depends again
on angle-variables. But introducing $\varepsilon=h^2$ as another small parameter, we can apply a KAM procedure,
under some non-resonance condition, to replace $(I,\varphi)$ by new symplectic coordinates $(I',\varphi')$ and thus get rid 
of angle dependance in the ${\cal O}(h^2)$ term, which yields $H(p,x,h)=H_0(I)+h^2\widetilde H_1(I)+\cdots$. 
See [CdV2], [Vu] for the case of commuting $h$-PDO's.

For a quasi-integrable system, we can combine the BNF (at the level of principal symbol) with the procedure above. 

In our case however, mostly because we content with the principal and subprincipal symbol of $H$, 
we have no real need to recover the canonical variables $(p,x)$ from $(I,\varphi)$, or $(J,\varphi)$,
nor to quantize action-angle variables.

\bigskip
\noindent {\bf 2. WKB solutions along perturbed tori}.
\smallskip
In this section, we carry out
WKB constructions associated to the family of perturbed tori $\Lambda^J$. We start from tori $\Lambda^J$ on which $H$ verifies
(1.12), and then extend the argument for tori obtained as in Theorem 1.2, which are
closer to be invariant by the hamiltonian flow. We use essentially known results on geometry of tori that we have
recalled in Appendix.
\medskip
\noindent {\bf a) Action integrals}.
\smallskip
We first compute the action integrals on the
lagrangian tori $\Lambda^J\subset T^*M$, constructed from $\Lambda$ with $J=I+\iota$, in terms of the action-angle variables $(J,\varphi)$.
We denote by ${\cal F}$ the set of focal points, that projects onto the caustics ${\cal C}$ (see Appendix A.)

To start with, consider $a,b\in M$ be away from ${\cal C}$. Assume $y=(p(b),b)\in\Lambda$ is not a focal point,
i.e. ${\partial X^0\over\partial\varphi}|_{\varphi=\varphi^I(b)}$
is not singular~; here $\varphi^I(b)$ is the angle-coordinate of $b$ on $\Lambda$. 
For $J\in\neigh (I)$, let $\iota_J:{\bf T}^d\to T^*M$ be the lagrangian embedding with image $\Lambda^J$
(the context prevents from confusing the map $\iota_J$ with the action variables $\iota$),
$d\phi^J=\iota_J^*(pdx)$ the 1-form parametrizing 
$\Lambda^J$ near $b$, and  $\varphi^J(x)$ the angle-coordinate
of $x\in\neigh (b)$ on $\Lambda^J$. 

Let $\gamma\subset\Lambda$ have its $x$-projection contained in the domain of definition of the $\phi^J$.
We want to compare $\int_\gamma pdx$ with 
$\int_{\gamma'} p'dx'$ on $\gamma'\subset\Lambda^J$, where $y'=(p',x')=Y(J,\varphi)$ are symplectic coordinates as in (1.5),
the $x$-projection of these paths joining $a$ to $b$. 
So $\gamma$ and $\gamma'$ are parametrized by $\varphi$, the action variable
being held constant. Along $\gamma$, $x=X^0(\varphi), 
p=P^0(\varphi)$, while along  $\gamma'$, $x'=X^0(\varphi)+X^J(\varphi)\iota
+{\cal O}(\iota^2)$, and 
$p'=P^0(\varphi)+P^J(\varphi)\iota+{\cal O}(\iota^2)$, with
$X^J(\varphi)$ and $P^J(\varphi)$ as in (1.2). Substituting these values into $\int_{\gamma'} p'dx'$ we get
$$\eqalign{
&\int_{\gamma'} p'dx'=\int_{\gamma'}\langle P^0(\varphi),dX^0(\varphi)\rangle
+\int_{\gamma'}d\langle P^0(\varphi),X^J(\varphi)\iota\rangle\cr
&+\int_{\gamma'}\langle P^J(\varphi)\iota,dX^0(\varphi)\rangle-
\langle dP^0(\varphi),X^J(\varphi)\iota\rangle+{\cal O}(\iota^2)\cr
}\leqno(2.1)$$
We choose the origin of angular coordinates such that $\varphi^J(a)=0$. Because $\Lambda^J$ is lagrangian, $\int_{\gamma'} p'dx'$
only depends on the end-points of $\gamma'$. We have
$$\int_{\gamma'}d\langle P^0(\varphi),X^J(\varphi)\iota\rangle=
\langle P^0(\varphi),X^J(\varphi)\iota\rangle|_0^{\varphi^J(b)}
\leqno(2.2)$$
Next we want to replace the path $\gamma'$ in 
$\int_{\gamma'}\langle P^0(\varphi),dX^0(\varphi)\rangle$ by
$\gamma$, which we complete into a  1-cycle in the 
$\varphi$-space as $\gamma'=\gamma'''+\gamma+\gamma''$, with
$\gamma'''$ between 0 and $\varphi^I(a)$, $\gamma''$ between $\varphi^I(b)$
and $\varphi^J(b)$. By Stokes' formula  and since $dP^0(\varphi)\wedge dX^0(\varphi)=0$~:
$$\int_{\gamma'}\langle P^0(\varphi),dX^0(\varphi)\rangle=
\bigl(\int_\gamma+\int_{\gamma''}+\int_{\gamma'''}\bigr)\langle 
P^0(\varphi),dX^0(\varphi)\rangle
\leqno(2.3)$$
We express the fact that $b=X(\varphi^J(b))=X^0(\varphi^I(b))$ as 
$$\langle \int_0^1{\partial X^0\over\partial\varphi}(\varphi(s))ds,
\varphi^J(b)-\varphi^I(b)\rangle=-X^J(\varphi^J(b))G(J,\varphi^J(b))\leqno(2.4)$$
where $\varphi(s)=\varphi^I(b)+s(\varphi^J(b)-\varphi^I(b))$ is a straight line. By continuity, $\varphi(s)\to\varphi^I(b)$
uniformly on $s\in[0,1]$ as $\iota\to 0$. 
Since $b$ is not a focal point on $\Lambda$, ${\partial X^0\over\partial\varphi}(\varphi^I(b))$ is not singular, and
the same holds for $\int_0^1{\partial X^0\over\partial\varphi}(\varphi(s))ds$ when 
$\iota$ is small enough, so (2.4) gives 
$\varphi^J(b)-\varphi^I(b)={\cal O}(\iota)$. 
Using also $\int_0^1{\partial X^0\over\partial\varphi}(\varphi(s))ds
-{\partial X^0\over\partial\varphi}(\varphi(t))={\cal O}(\iota)$
uniformly on $t\in[0,1]$ and 
$\int_0^1P^0(\varphi(t))dt-P^0(\varphi^J(b))={\cal O}(\iota)$
we find
$$\int_{\gamma''}\langle P^0(\varphi),dX^0(\varphi)\rangle
+\langle P^0(\varphi),X^J(\varphi)\iota\rangle|_{\varphi=\varphi^J(b)}={\cal O}(\iota^2)$$
The same holds for the contribution of $\gamma'''$ since we may
always assume, $a$ being fixed, that it is not a focal point, so these
estimates shows with (2.2) that 
$$\int_{\gamma'}d\langle P^0(\varphi),X^J(\varphi)\iota\rangle
+\bigl(\int_{\gamma''}+\int_{\gamma'''}\bigr)\langle 
P^0(\varphi),dX^0(\varphi)\rangle={\cal O}(\iota^2)\leqno(2.5)$$
On the other  hand, using the canonical relations (1.7) for 
coordinates $y'$, and the estimate $G(J,\varphi)=\iota+{\cal O}(\iota^2)$ in $C^1$-topology, one readily verifies that 
$$\langle P^J(\varphi)\iota,dX^0(\varphi)\rangle-
\langle dP^0(\varphi),X^J(\varphi)\iota\rangle=\langle\iota+{\cal O}(\iota^2),d\varphi\rangle$$
Integrating this 1-form along $\gamma'$, where $\iota$ is constant, 
and inserting in (2.1) together with (2.5) we have~:
$$\int_{\gamma'}p'dx'=\int_{\gamma}pdx+\langle\varphi^J(b)-\varphi^J(a),\iota\rangle+{\cal O}(\iota^2)\leqno(2.6)$$
More generally, for $N=2,3\cdots$, $\kappa_N$ as in Theorem 1.2 and $J'\in\neigh (I)$, consider the lagrangian embedding
$\iota_{J',N}=\iota\circ\kappa_N:{\bf T}^d\to T^*M$, with image $\Lambda^{J',N}$. Let again 
$d\phi^{J',N}=\iota_{J',N}^*(pdx)$ parametrize $\Lambda^{J',N}$ near $b$. Using the estimates $\varphi'=\varphi+{\cal O}(\iota'^2)$,
$J=J'+{\cal O}(\iota'^2)$, one can readily extend (2.6) to the tori $\Lambda^{J',N}$, so we get~:
\smallskip
\noindent {\bf Proposition 2.1}:  Let $a$ be away from the caustics of 
$\Lambda$ and $y=(p(b),b)\in\Lambda$ is not a focal point, and assume $a$ and $b$ can be joined by a path
$\gamma\subset\Lambda$ contained in the domain where $\Lambda$ is projectable
(i.e. the $x$-projection of $\gamma$ is contained in the domain of definition of $\phi^I$, 
where  the 1-form $d\phi^I=\iota_I^*(pdx)$ is a local parametrisation of 
$\Lambda$.~)
Then for any $N=1,2,\cdots$, and $\iota'$ small enough, for all tori $\Lambda^{J',N}$ as above, we have~: 
$$\phi^{J',N}(b)-\phi^{J',N}(a)=\phi^I(b)-\phi^I(a)+\langle\varphi^{J',N}(b)-\varphi^{J',N}(a),\iota'\rangle
+{\cal O}(\iota'^2)\leqno(2.7)$$
\medskip
Note that (2.7) can be written at the infinitesimal level as
$\iota_{J',N}^*(pdx)=\iota_I^*(pdx)+\iota' d\varphi+{\cal O}(\iota'^2)$.

Next we consider the case where $y=(p(b),b)\in\Lambda$ is a focal point, 
but $y\notin {\cal F}_0$ [which we recall from Appendix A], so that  
${\partial P^0\over\partial\varphi}|_{\varphi=\varphi^I(b)}$,
(i.e. $d\pi_p:\Lambda\to{\bf R}^d_p$) is non singular. 
For small $\iota$, the ``alternate'' action is now $\int_{\widetilde
\gamma} -xdp$ and the new generating function $d\psi^I=\iota_I^*(-xdp)$. 
At points $y$ where $d\pi^I_x$ and $d\pi^I_p$ are non singular 
(i.e. where $\Lambda^I$ is bi-projectable) $\psi^I$ and $\phi^I$ are related by Legendre
transformation $\phi^I(x)=\langle -p,\nabla\psi^I(p)\rangle+\psi^I(p)|_{p=p(x)}$, so 
$\phi^I=\psi^I+\iota_I^*(xp)$, up to a constant.
We pick again $a$ away from the caustics,
and write $y=(\beta,x(\beta))$, $(p(a),a)=(\alpha,x(\alpha))$. For sufficiently small $\iota$ this holds for the neighborhing tori
$\Lambda^J$ and also for the $\Lambda^{J',N}$.

So we need just repeat the former argument, comparing 
$\int_{\widetilde\gamma'} -x'dp'$ with $\int_{\widetilde\gamma} -xdp$,
where the first integral is computed along a path $\widetilde\gamma'$ 
joining (the lift of) $\alpha$ with  $\beta$ on $\Lambda^J$, and the second  one
along a path $\widetilde\gamma$ 
joining $\alpha$ with $\beta$ on $\Lambda$, and contained in the domain of definition of $\psi^I$. 
Thus we get~:
\smallskip
\noindent {\bf Proposition 2.2}:  Assume $y=(p(b),b)=(\beta,x(\beta))
\in\Lambda$ is a focal point, but $y\notin {\cal F}_0$. If $a$ and $b$ can be joined by a path 
contained in the domain of definition of $\psi^I$, then 
for any $N=1,2,\cdots$, and $\iota'$ small enough, for all tori $\Lambda^{J',N}$ as above we have~: 
$$\psi^{J',N}(\beta)-\psi^{J',N}(\alpha)=\psi^I(\beta)-\psi^I(\alpha)+\langle
\varphi^{J',N}(b)-\varphi^{J',N}(a),\iota'\rangle+{\cal O}(\iota'^2)\leqno(2.8)$$
with notations as in Proposition 2.1, and where $d\psi^{J',N}=\iota_{J',N}^*(-xdp)$ is the 1-form par- ametrizing 
$\Lambda^{J',N}$ near $b$. 
\medskip
Note also that (2.8) can be written at the infinitesimal level as
$\iota_{J',N}^*(-xdp)=\iota_I^*(-xdp)+\iota' d\varphi+{\cal O}(\iota'^2)$.

Then we consider the general case, to take in account the possible exceptional set ${\cal F}_0$ defined in Appendix A so we
assume that $d\pi|_\Lambda(y)$ is of rank $k\leq d-1$. 
So $\Lambda$ can be parametrized by a Morse
function $S(x,\theta)$ defined near $\iota_I^{-1}(y)$, and  
by the discussion leading to (b.27), we may assume $S(x,\theta)$ is of the form 
$T(x_1,p_2)+\langle x_2,p_2\rangle$ and $\Lambda=\Lambda_T$. 

Take a base point $A=(\alpha_2,a_1)\in{\bf R}^k\times{\bf R}^{d-k}$
that we identify with its lift through $\widetilde\pi:\Lambda\to{\bf R}^d, (p,x)\mapsto(p_2,x_1)$
with $A\in\Lambda_T$, we assume that $A$  is non singular.
The variable point near $y$, with the same abuse of notations is now $B=(\beta_2,b_1)\in\Lambda_T$. 
We deform $\Lambda$ into $\Lambda^J$, and the generating function $T$ to $T^J$. We shall always
work in the domains  where $T^J$ are well defined, and drop the subscript $T^J$ from $\Lambda^J_{T^J}$. 
So let $\gamma\subset\Lambda$ be a path between (the lift of) $A$ and $B$, $\gamma=\gamma_1+\gamma_2$ such that 
$p_2=\beta_2$ along $\gamma_1$, and $x_1=a_1$ along $\gamma_2$. Using (b.27), we find 
$\int_{\gamma_1}p_1dx_1=\int^{x_1}{\partial  T\over\partial x_1}(x_1,p_2)dx_1|_{p_2=\beta_2}$ and
$\int_{\gamma_2}-x_2dp_2=\int^{p_2}{\partial  T\over\partial p_2}(x_1,p_2)dp_2|_{x_1=a_1}$, so
$$(T(x_1,p_2)+\langle  x_2,p_2\rangle)|_A^B=\int_{\gamma_1}p_1dx_1+\int_{\gamma_2}-x_2dp_2\leqno(2.10)$$
In the same way, we let $\gamma'\subset\Lambda^J$ be a path between $A$ and $B$, $\gamma'=\gamma'_1+\gamma'_2$ such that 
$p_2=\beta_2$ along $\gamma'_1$, and $x_1=a_1$ along $\gamma'_2$, so denoting $T'$ for $T^J$ we have~:
$$T'(x'_1,p'_2)+\langle  x'_2,p'_2\rangle|_A^B=\int_{\gamma'_1}p'_1dx'_1+\int_{\gamma'_2}-x'_2dp'_2\leqno(2.11)$$
We parametrize $\Lambda^I$ and $\Lambda^J$ by $\varphi$, so that $x=(x_1,x_2)=X^0(\varphi)=(X^0_1(\varphi),(X^0_2(\varphi))$,
along $\gamma$, and $x'=(x'_1,x'_2)=X(\varphi)=(X_1(\varphi),(X_2(\varphi))$,  along $\gamma'$, where
as above, $X(\varphi)=X^0(\varphi)+X^J(\varphi)\iota+{\cal O}(\iota^2)$, and 
$P(\varphi)=P^0(\varphi)+P^J(\varphi)\iota+{\cal O}(\iota^2)$. Possibly after renumbering the variables
$\varphi$, we may assume that we have a partition of the coordinates $\varphi=(\varphi_1,\varphi_2)$ 
adapted to the projection $\widetilde \pi:\Lambda\to{\bf R}^k\times{\bf R}^{d-k}$, i.e.  such that
${\partial  X^0_1\over\partial\varphi_1}$ and ${\partial  P^0_2\over\partial\varphi_2}$
are non singular. Then the argument of Propositions 2.1 and 2.2 carries over to the present situation, 
working separately on $\gamma_1$ and $\gamma'_1$ with the $x_1$ and $x'_1$ coordinates, then on $\gamma_2$ and $\gamma'_2$
with the $p_2$ and  $p'_2$ coordinates. So we proved the following generalization of Propositions 2.1 and 2.2~:
\medskip
\noindent {\bf Proposition 2.3}:  Assume $y=(p(b),b)\in\Lambda$ is a general focal point, let
$\Lambda=\Lambda^I$ is parametrized as in (b.27), and pick a non singular poiint $A\in\Lambda$ near $y$. 
Then for $\iota$ small enough and $B$ near $y$, 
$$\bigl(T^J(x'_1,p'_2)+\langle x'_2,p'_2\rangle\bigr)|_A^B=\bigl(T^I(x_1,p_2)+\langle
x_2,p_2\rangle\bigr)|_A^B+\langle\varphi^J(b)-\varphi^J(a),\iota\rangle+{\cal O}(\iota^2)\leqno(2.12)$$
Here $d(T^J(x'_1,p'_2)+\langle x'_2,p'_2\rangle)=\iota_J^*(-xdp)=\iota_J^*(pdx)$ is the 1-form parametrizing 
$\Lambda_J$ near $B$ where it is bi-projectable, and  $\varphi^J(B)\in{\bf T}^d$  the angle coordinate of $B$.
More generally, let 
$N=2,3,\cdots$, and $\iota'$ small enough, for all tori $\Lambda^{J',N}$ as above we have
with obvious notations~: 
$$\bigl(T^{J',N}(x'_1,p'_2)+\langle x'_2,p'_2\rangle\bigr)|_A^B=\bigl(T^I(x_1,p_2)+\langle
x_2,p_2\rangle\bigr)|_A^B+\langle\varphi^{J',N}(b)-\varphi^{J',N}(a),\iota'\rangle+{\cal O}(\iota'^2)\leqno(2.13)$$
\medskip
\noindent{\bf b) WKB solutions}.
\smallskip
The constructions are recalled for completeness in Appendix A in case of a Schr\"odinger operator $-h^2\Delta+V(x)$, but the result
holds for more general $h$-PDO. The local WKB solution near point $x_0\in M$ depends again on the structure of $\Lambda$ above $x_0$.

First we describe locally the quasi-mode associated with $\Lambda^I$ in term of $(P^0(\varphi),X^0(\varphi))$.
In the simplest case described in Proposition 2.1, i.e. where $\Lambda$ is projectable,
the phase function restricted to $\Lambda$ equals $\phi^I(\varphi)$, it verifies $P^0(\varphi)=\nabla\phi^I(\varphi)$
and solves Hamilton-Jacobi equation $H(\nabla\phi^I(\varphi),X^0(\varphi))=H_0=E$. Recall from (b.4) that
the solution to the transport equation verifies
$a_0(x;E)|_{x=X^0(\varphi)}=\Const |\det {\partial X^0\over\partial\varphi}|^{-1/2}$.

Denote for short by $\int^x pdx$ a solution of Hamilton-Jacobi equation with data on $\Lambda$, and 
let $u(x;E,h)=a_0(x;E)e^{i(\int^x pdx)/h}$. Then $u$ solves $(H(x,hD_x)-E)u={\cal O}(h^2)$.
Here $H(x,hD_x)$ denotes any ``reasonable'' $h$-quantization of $H(p,x)$, e.g. Weyl quantization~; note that as usual, we first quantize  
the variables $(p,x)$, then restrict to the torus. Moreover
$u(x;E,h)|_{x=X^0(\varphi)}=\Const |\det {\partial X^0\over\partial\varphi}|^{-1/2}e^{i\phi^I(X^0(\varphi))/h}$.

For $N=2,3\cdots$ let us then consider, for small $\iota'$ 
$$u(x;\iota',h)|_{x=X(J',\varphi')}=\Const |\det {\partial X(J',\varphi')\over\partial\varphi'}|^{-1/2}
e^{i\phi^{J',N}\bigl(X(J',\varphi')\bigr)/h}, 
\ \varphi'\in{\bf T}^d\leqno(2.15)$$
Clearly, $u(x;\iota',h)|_{x=X(J',\varphi')}$ 
is a lagrangian distribution supported on $\Lambda^{J',N}$, and 
$$\bigl[H(x,hD_x)u(x;\iota',h)\bigr]|_{x=X(J',\varphi')}$$ 
is obtained by restricting $H(x,hD_x)\bigl[a_0(x;E)e^{i(\int^x pdx)/h}\bigr]$
to $x=X(J',\varphi')$, and $p=P(J',\varphi')$. Let also $E_N=H_N(\iota')$. At the level of the eikonal equation we have by (1.15)
$$(H(p,x)-E_N)|_{y=(P(J',\varphi'),X(J',\varphi'))}={\cal O}(\iota'^{N+1})\leqno(2.16)$$
Now we consider the transport equation and compute, according to (b.3), (b.4)
$${\cal L}_{X_H|_{\Lambda^{J',N}}}(a_0(x,E)|dx|^{1/2})={\cal L}_{X_H|_{\Lambda^{J',N}}}(|d\varphi|^{1/2})\leqno(2.17)$$
Here ${\cal L}_{X_H}$ denotes the Lie derivative of Hamilton vector field $X_H$ acting on
half-densities, $X_H$ is no longer tangent to $\Lambda^{J',N}$, but it follows again from (1.15) that  
${\cal L}_{X_H|_{\Lambda^{J,N}}}(|d\varphi|^{1/2})={\cal O}(\iota'^{N+1})$.
So if we choose $N$ such that $\iota'^{N+1}={\cal O} (h^2)$, we get
$$\bigl[(H(x,hD_x)-E_N)u(x;\iota',h)\bigr]|_{x=X(J',\varphi')}={\cal O}(h^2)\leqno(2.18)$$ 
Now we turn to the case described in Proposition 2.2, i.e. 
near a focal point $y_0=(p_0,x_0)$ where $\Lambda$ is $\pi_p$-projectable.
The phase function restricted to $\Lambda$ equals $\psi^I(\varphi)$, it verifies $X^0(\varphi)=\nabla\psi^I(\varphi)$
and solves Hamilton-Jacobi equation $H(\nabla\psi^I(\varphi),P^0(\varphi))=H_0=E$. 
The quasi-mode restricted to $\Lambda$ near $x_0$ is then given by (b.5), with phase $\psi^I$, 
and principal symbol $b_0(p,E)=C|\det {\partial  P^0\over\partial \varphi}|^{-1/2}$ that verifies (b.15). 

So for $N=2,3,\cdots$, and $\iota'$ small enough, let 
$\chi\in C_0^\infty$ is a cut-off equal to 1 near the angle-coordinate of $p_0$,
substitute as above $(P(J',\varphi'),X(J',\varphi'))$ for $(p,x)$ in (b.5), and define
$$v(x;\iota',h)|_{x=X(J',\varphi')}=(2\pi h)^{-d/2}\int_{{\bf T}^d}
e^{i\bigl(\langle x,P(J',\theta)\rangle+\psi^{J',N}(P(J',\theta))\bigr)/h}\chi(\theta)
|\det {\partial P(J',\theta)\over\partial\theta}|^{1/2}d\theta \leqno(2.19)$$
Following the proof of Theorem b.1 leading to (b.15) we see that $v(x;\iota',h)|_{x=X(J',\varphi')}$ also verifies (2.18), with
the same energy $E_N=H_N(\iota')$ that depends only on $\Lambda^{J',N}$. 

At last, we can consider the general focal point $y_0$ described in Proposition 2.3, and
use the general representation of the quasi-mode supported on $\Lambda$ in a neighborhood of $x_0$ given by (b.17), i.e.
$w(x;E,h)=(2\pi h)^{-d/2}\int e^{iS(x,\theta)/h}a(x,\theta,E,h)dx$, where we have chosen the number of phase variables
equal to $d$. It is suitable to express $w(x;E,h)$ in the mixed representation described after Theorem b.1, so take
$(x_1,p_2)$ as local coordinates on $\Lambda$,  $\varphi=\varphi(x_1,p_2)$. For the unperturbed torus $\Lambda$, we have 
$$\eqalign{
w(x,h)&|_{x=X(\varphi)}=(2\pi h)^{-d/2}\int e^{i\langle x_2,p_2\rangle/h}\cr
&\times\exp [i(\int^y\langle P,dX\rangle-\langle P_2(x_1,p_2),X_2(x_1,p_2)\rangle)/h]\big|_{y=(P(x_1,p_2),X(x_1,p_2))}\cr
&\times\chi\bigl(\varphi(x_1,p_2)\bigr)
|\det {\partial (X_1(\varphi), P_2(\varphi))\over\partial\varphi}|^{-1/2}dp_2 
\cr}\leqno(2.20)$$
Then we argue as before, replacing $(I,\varphi)$ by $(J',\varphi')$ in (2.20), and verify it satisfies (2.18) when 
$\iota'^{N+1}={\cal O} (h^2)$. To get a quasi-mode we need to glue different lagrangian distributions expressed locally as
$u,v,w$ and take care of Maslov indices as is explained in Appendix C. There we notice that the embedding $(\Lambda,\iota)$ 
is topologically
stable if we exclude some exceptional lagrangian singularities in $\Lambda$, and so Maslov indices for the $\Lambda^{J',N}$
remain the same. 

On the other hand, Propositions 2.1-3 and implicit functions theorem 
show that, given $\delta>0$, it is indeed possible to ``match'' $\iota'$ with values of the actions and Maslov 
indices $\alpha$ along the fondamental cycles as in (1.10) so to satisfy Maslov quantization condition ${1\over h}\cdot J'+\alpha/4=0$
mod $H^1(\Lambda;{\bf Z})=H^1(\Lambda^{J',N};{\bf Z})$, for a sequence 
$J'=J'_k(h)$. So let us summarize our main result in the
\medskip
\noindent {\bf Theorem 2.4}: Let $(\Lambda,\iota)$ as before such that $\Lambda$ is invariant under the Hamiltonian flow of $H$
and this flow has diophantine frequencies $\omega$, $H|_{\Lambda}=H_0=E$. Assume $\Lambda$ is free of exceptional lagrangian singularities 
so that $H^1(\Lambda;{\bf R})$ is stable under small perturbations. Let also $\delta>0$. Then there is a sequence of tori $\Lambda^{J',N}$,
for $J'=J'_k(h)$ satisfying $|k|h\leq h^\delta$, 
such that $H$ has quasi-modes $(u_k(x,h),E_k(h)), |E_k(h)-E|={\cal O}(h^\delta)$, in the sense
$(H(x,hD_x)-E_k(h))u_k(x,h)={\cal O}(h^2)$. Morover, these quasi-modes are of maximal density, in a $h^\delta$-neighborhood of 
$\Lambda$. 
\smallskip
\noindent {\it Remark 2.5}: 
In the case $H$ contains a subprincipal symbol and equation (b.29) holds,
the sub-principal form makes sense on $\Lambda$, only when it is invariant by $X_H$. 
But again, the diophantine condition and our special action-angle variables 
allow to extend these constructions to the case where $X_H$ is no longer tangent to $\Lambda$, 
provided we solve (b.29) as in (b.31)
instead of (b.25), and modify Maslov quantization condition as in (c.4).

\bigskip
\noindent {\bf 3. Jacobi-Maupertuis correspondance}.
\medskip
We will apply the constructions above to Hamiltonian $H(p,x)=p^2+V(x)$ near energy $E=0$, where $V(x)=-{\cal E}/g(x)$. 
The ``reference Hamiltonian''  ${\cal H}=g(x)p^2$ denotes an 
(covariant) metric on $T^*M$, for which $\Lambda\subset\{{\cal H}={\cal E}\}$ is an integral lagrangian torus with 
irrationnal frequencies $\widetilde\omega$.
The Hamilton vector fields are related by $X_{\cal H}=g(x)X_H$, so that if $\tau$ parametrizes the integral curves of
$X_{\cal H}$ and $t$ (the new time) those of $X_H$, we have the relation 
$dt=g(X^0(\tau))d\tau$. This is Maupertuis-Jacobi correspondance.
\medskip
\noindent {\bf a) Construction of quasi-modes}.
\smallskip
First we show that,
assuming a diophantine condition on $\widetilde\omega$, the flow on $\Lambda$ for $H$ is again quasi-periodic.  
Expand $g\in C^\infty$ as a Fourier  series on 
$\Lambda\approx{\bf R}^d/2\pi{\bf Z}^d$,
$g\bigl(X^0(\varphi)\bigr)=\langle g\rangle+\Sum_{n\in{\bf Z}^d\setminus 0}c_n e^{in\varphi}$, $c_n$ rapidly decreasing, we have~:
\medskip
\noindent {\bf Proposition 3.1}: Assume a diophantine condition on $\widetilde\omega$. Then Maupertuis-Jacobi correspondance 
induces a reparametrization of $\Lambda$ by 
$$\Phi : \varphi_0\mapsto\varphi=\Phi(\varphi_0)=\varphi_0+\widetilde\omega f(\varphi_0)$$ 
where $f$ is a smooth, scalar periodic function
with $\langle f\rangle=0$. The motion on $\Lambda$ induced by this reparametrization is quasi-periodic with frequency vector
$\omega=\widetilde\omega/\langle g\rangle$, and (for simplicity we denoted $g\circ X^0$ by $g$)~:
$$\det {\partial\Phi\over\partial\varphi_0}(\varphi_0)={\langle g\rangle \over g}\leqno(3.1)$$ 
\smallskip
\noindent {\it Proof}: Integrating $dt=g(\widetilde\omega \tau+\varphi_0)d\tau$  we get
$$t=t_0+\langle g\rangle\tau+F(\widetilde\omega \tau+\varphi_0)-F(\varphi_0)$$
where $F$ is obtained by integrating
$g$ term by term, and $t_0$ is an arbitrary constant. We have $\langle F\rangle=0$, and $F$ is again
smooth due to the diophantine condition. Choose $t_0=F(\varphi_0)$, and change variable $z=\tau-t/\langle g\rangle$.
With the new frequency $\omega=\widetilde\omega/\langle g\rangle$, the latter equation rewrites as
$$h(z,\psi_t)=z+F(\psi_t+\widetilde\omega z)/\langle g\rangle=0\leqno(3.2)$$
where we have set $\psi_t=\varphi_0+\omega t$. Equation (3.1) holds when $z=z_0=-t_0/\langle g\rangle$,
$\psi_t=\psi_{t_0}=\varphi_0+\omega t_0$, and moreover ${\partial h\over\partial z}=g(\psi_t+\widetilde\omega z)
/\langle g\rangle>0$. So implicit function theorem shows that the solution to (3.2) is of the form
$z=f(\psi_t)$, $\psi_t\in\Lambda$, where $f$ is periodic and $\langle f\rangle=0$. 
So far we have shown that Maupertuis-Jacobi correspondance can be written as 
$\varphi_0+\widetilde\omega\tau=\varphi_0+\omega t+\omega f(\varphi_0+\omega t)$. To get the Proposition, it suffices to
observe that the trajectory $t\mapsto\varphi_0+\omega t$ is everywhere
dense in $\Lambda$, since $\omega$ is again irrationnal, and moreover 
${\partial\over\partial\varphi}(\varphi+\widetilde\omega f(\varphi)=I+\widetilde\omega f'(\varphi)$
is of maximal rank, because
$$\det \bigl(I+\widetilde\omega f'(\varphi)\bigr)={\langle g\rangle \over g}$$
which brings the proof to an end. $\clubsuit$
\smallskip
Note that $\omega$ and $\widetilde\omega$ are simultanously diophantine.
We can now construct quasi-modes for $H$ supported near $\Lambda$ using the procedure set up in Sect.2. 
On $\Lambda$, Maupertuis-Jacobi correspondance will only induce by (3.2) a change in the densities, namely
$$\det {\partial X^0\over\partial\varphi}={\langle g\rangle\over g}\det {\partial X^0\over\partial\varphi_0}, \
\det {\partial  P^0\over\partial \varphi}={\langle g\rangle\over g}\det 
{\partial  P^0\over\partial \varphi_0}\leqno(3.3)$$
and also in the mixed representation. 

On neighboring tori $\Lambda^{J',N}$, we have $(J',\varphi')=\kappa_N^{-1}(J,\varphi)=\kappa_N^{-1}(J,\Phi(\varphi_0))$,
so that we can still consider $\varphi'$, for given $J$, as a function of $\varphi_0$ alone. 
Let us summarize our result so far in the~:
\medskip
\noindent {\bf Theorem 3.2}: Let ${\cal H}=g(x)p^2$ be a (not necessarily integrable) covariant metric on $T^*M$
have an invariant torus 
$\Lambda$ with ${\cal H}|_{\Lambda}={\cal E}$, and $I$ be the actions 
as in (1.1). Assume that on $\Lambda$ the Hamiltonian flow of ${\cal H}$ 
has diophantine frequencies $\widetilde\omega$,  
and morerover $\Lambda$ is free of exceptional lagrangian singularities 
so that $H^1(\Lambda;{\bf R})$ is stable under small perturbations. 

On the other hand let $H(p,x)=p^2+V(x)$, where $V(x)=-{\cal E}/g(x)$, be the  hamiltonian on $T^*M$
given by Maupertuis-Jacobi correspondaance.
Let also $\delta>0$. Then there is a sequence of tori $\Lambda^{J'}$,
for $J'=J'_k(h)$ satisfying $|k|h\leq h^\delta$, 
such that $H$ near energy $E=0$ has quasi-modes $(u_k(x,h),E_k(h)), |E_k(h)-E|={\cal O}(h^\delta)$, 
supported on $\Lambda^{J',N}$, in the sense
$(H(x,hD_x)-E_k(h))u_k(x,h)={\cal O}(h^2)$. Morover, these quasi-modes are of maximal density, in a $h^\delta$-neighborhood of 
$\Lambda$. The quasi-eigenfunctions $u(x,\iota',h)=u_k(x,h)$ 
can be expressed in term of ``Maslov canonical operator'', with the help of formulas (2.15), (2.19), (2.20).
In particular where $\Lambda$ is projectable, we have~:
$$u(x;\iota',h)|_{x=X(J',\varphi')}=\Const |\det {\partial X(J',\varphi')\over\partial\varphi'}|^{-1/2}
e^{i\phi^{J',N}\bigl(X(J',\varphi')\bigr)/h}|_{(J',\varphi')=\kappa_N^{-1}(J,\Phi(\varphi_0))}\leqno(3.4)$$ 
where $\varphi_0\in{\bf T}^d$. So the half-density $|\det {\partial X(J',\varphi')\over\partial\varphi'}|^{-1/2}$, for given $J'$,
can be written as a function of $\varphi_0$ alone. 
\smallskip
Instead of $H(x,hD_x)=-\Delta^2+V(x)$ we may consider a $h$-PDO with symbol 
$H(p,x;h)=p^2+V(x)+hH_1(p,x)+\cdots$. The introduction of a sub-principal symbol can be handled similarly, due to Remark 2.5. It
modifies $|\det {\partial X(J',\varphi')\over\partial\varphi'}|^{-1/2}$ by a factor 
$$\exp [-iG(\varphi')]|_{(J',\varphi')=\kappa_N^{-1}(J,\Phi(\varphi_0))}$$ 
as in (b.31), and Maslov index by the vector
$\langle\sigma_H\rangle|_{\Lambda^{J',N}}(1,\cdots,1)$ as in (c.4).

More generally, this procedure may be applied whenever Maupertuis-Jacobi correspondance relates the flows of
two classical Hamiltonians 
${\cal H}$ and $H$ on a lagrangian manifold $\Lambda$, $H$ not necessarily of the form $p^2-1/g(x)$, as is the 
case for the shallow water waves dispersion relation of Sect.4.d. Actually, all what we need is to construct an 
oscillating principal symbol for lagrangian distributions microlocalized on the family $\Lambda^{J',N}$, and this can be done
following the standard procedure we recall in Appendix b. 

\medskip
\noindent {\bf b) Remarks on isospectrality}.
\smallskip
There naturally arises the question of comparing the quasi-modes microlocalized near 
$\Lambda$ of the Schr\"odinger operator 
$H=-h^2\Delta-E/g(x)$, as given in Theorem 3.2, with those of the Laplacian ${\cal H}(x,hD_x)hD_xg(x)hD_x$ on $M$, in other words~:
does Jacobi-Maupertuis correspondance imply (partial) isospectrality ? Theorem 3.2 gives all quasi-modes of $H$ supported in a 
microlocal neighborhood of an invariant torus $\Lambda$ for ${\cal H}$. Here we address the converse problem, i.e. how
to construct quasi-modes for the Laplace operator associated with ${\cal H}$ (that we may write in a symmetric form as 
${\cal H}(x,hD)=\langle hD_x,g(x)hD_x\rangle$ see e.g. [Mat],~) with quasi-eigenvalues close to the $E_k(h)$ ?  

We shall assume here that ${\cal H}=g(x)p^2$ is integrable, so that in global action-angle variables, 
${\cal H}={\cal H}(I)=\langle \widetilde\omega(I),I\rangle$, where $I$ are as in (1.1), 
and $ \widetilde\omega(I)={\partial {\cal H}\over\partial I}$.
Choose $I=I_0$ as a base point for the action-angle coordinates $(J',\varphi')$ we have constructed so far for $H$, and consider
${\cal H}(J')=\langle \widetilde\omega(J'),J'\rangle$.  Under some generic non degeneracy condition, the integral manifolds
of ${\cal H}(J')$ are close to those of ${\cal H}(I)$, for small enough $\iota'$. 
By definition, the action integrals on the lagrangian tori $\widetilde\Lambda^{J'}$ 
(integral manifolds for ${\cal H}$) verify (1.1), and the
$\widetilde\Lambda^{J'}$ are close enough to the $\Lambda^{J',N}$ (integral manifolds for $H$), although in general
$\widetilde\Lambda^{J'}\neq\Lambda^{J',N}$ if $\iota'\neq0$. It is still possible to compare the action $\int pdx$ along $\Lambda$ 
with the action $\int pdx$ along $\widetilde\Lambda^{J'}$ as in Propositions 2.1-3. We can also extend the WKB constructions leading
to Theorem 2.4 in the case of the Schr\"odinger operator $H$ to the Laplace operator ${\cal H}(x,hD)$,
and the fact that $X_{\cal H}$ is tangent to $\widetilde\Lambda^{J'}$ makes them easier. To get Maslov quantization condition on 
$\widetilde\Lambda^{J'}$ it suffices to choose   
$J'=\widetilde J'_k(h)$  (instead of $J'=J'_k(h)$). Since $\widetilde J'_k(h)=J'_k(h)(1+{\cal O}(h))$, 
this gives easily the required (approximate) isospectrality result. 

\bigskip
\noindent {\bf 4. Some applications}
\medskip
We start by examining the simple case of a conformal, radially
symmetric metric. Then both geodesic flow and  hamiltonian flow (with  potential) are
integrable, and the constructions above are not truly necessary, but this
can  help to understand  more complicated situations, which we
consider next. Finally we study the Hamiltonian
associated with the dispersion relation that arises in the shallow water waves theory.
\medskip
\noindent {\bf a) Radial potential and related Zeeman effect}.
\smallskip
The metric in cotangent space $T^*{\bf R}^2$ we consider is given by ${\cal H}(p,x)=p^2g(|x|)$.
In polar coordinates $x=(r,\theta)$ with dual variables $p=(\rho,\psi)$, 
the kinetic energy reads $p^2=\rho^2+{\psi^2\over  r^2}$ and
Hamilton equations take the form
$$\eqalign{
&\dot r=2\rho g(r), \quad 
\dot \rho=-g'(r)\bigl(\rho^2+{\psi^2\over r^2}\bigr)+2{g(r)\over r^3}\psi^2\cr 
&\dot\theta=2\psi{g(r)\over r^2}, \quad \dot\psi=0\cr
}\leqno(4.1)$$
(Because of the radial symmetry, this is equivalent to consider the Hamiltonian with potential.)
This example is of course well known (see [Ze], [HiSjVu] and references therein)
but we want here to give an explicit parametrization for $(r,\theta)$ in action-angle variables, from which we 
deduce the configuration of caustics.
We assume the energy surface $\Sigma_{\cal E}$
$$g(r)\bigl(\rho^2+{\psi^2\over r^2}\bigr)={\cal E}\leqno(4.2)$$
to be compact, and look for the lagrangian foliation of $\Sigma_{\cal E}$. The last
equation (4.1) shows that $\psi=\psi_0$ is a constant of the
motion, so the moment map is of the form
$m=\bigl(g(r)\bigl(\rho^2+{\psi^2\over r^2}\bigr),\psi)$.
Let us find the actions variables and the fundamental frequencies $\omega=(\omega_1,\omega_2)$
on the torus $\Lambda=\{m=({\cal E}, \psi_0)\}$. We have 
$$\langle P,dX\rangle=\rho dr+\psi d\theta\leqno(4.3)$$
Assume there is $r_0>0$ such that $\psi_0^2{g(r_0)\over r_0^2}={\cal E}$. 
Then ${\cal C}=\{r=r_0\}$ is a caustic of $\Lambda$ (turning surface)  and for
$r=r_0$, $\rho=0$. 
For a general Morse function $g(r)$, there are several turning surfaces, 
delimiting annuli of radius $r_1<\cdots<r_n$ in the $x$-space. Choose an annulus $D=\{r_1<r<r_2\}$
in the classically allowed region, and call $\Lambda$ the connected component 
of the integral manifold lying above $D$. 
As a first fundamental cycle $\gamma_1$ in
$\Lambda$, we take $r=r_0$, ($r_0=r_1$ or $r_2$), $\psi=\psi_0$. Integrating
(4.3) along $\gamma_1$, we find the first action variable
$I_1={1\over 2\pi}\int_{\gamma_1}\langle P,dx\rangle=\psi_0$ (angular momentum). 
Let $\gamma_2$ the second  fundamental cycle be given by $\theta=\theta_0$, $r_1\leq r\leq r_2$, this yields
Clairaut integral
$$I_2={1\over 2\pi}\int_{\gamma_2}\langle P,dx\rangle={1\over 2\pi}\int_{r_1}^{r_2}\rho dr=
\int_{r_1}^{r_2}\bigl({{\cal E}\over g(r)}-{\psi_0^2\over r^2}\bigr)^{1/2}dr$$
Substituting $\psi_0=I_1$, since the turning points $r_1,r_2$ depend 
also on $I_1$, we find that $I_2$ is of the form $I_2=f(I_1)$ where $f$ is a smooth function (recall that ${\cal E}$
is held fixed.) The hamiltonian reduces to ${\cal H}(I_1,f(I_1))={\cal E}$, and differentiating this relation
we get the rotation (or winding) number 
$${\omega_1\over\omega_2}(\psi_0)={\partial_{I_1}{\cal H}\over\partial_{I_2}{\cal H}}=-f'(I_1)$$
Next we notice that the 2 first equations (4.1) relative to the $(r,\rho)$-projection of motion are decoupled from
the other equations for the $\theta$-projection.
So integrating  these
and using that $\Lambda$ is a torus of periods $2\pi$, we find that 
$$r=r(\omega_2 t+\varphi_2^0), \quad \rho=\rho(\omega_2 t+\varphi_2^0)\leqno(4.4)$$
for some constant $\varphi_2^0$.
Now we use the third equation (4.1) which can be integrated as
$\theta(t)=2\psi_0\int^t{g(r)\over r^2}|_{r=r(\omega_2s+\varphi_2^0)}ds$.
We expand ${g(r)\over r^2}$ as a Fourier series in $\omega_2s+\varphi_2^0$, and integrate
term by term, which gives 
$$\theta=\widetilde\theta(\omega_2t+\varphi_2^0)+\omega_1t+\varphi_1^0\leqno(4.5)$$
Here $\varphi_1^0$ is a constant, and $\omega_1=2\psi_0\langle{g(r)\over r^2}\rangle$, the 0:th order Fourier
coefficient, ($\langle\cdot\rangle$ denotes the time average over $[0,2\pi]$.)

We look now at the caustic and determine the focal points. Note that because of conservation of momentum, to any $x$ inside the annulus, 
correspond exactly 2 values of $p$. We have
$$ C={\partial X\over\partial
  \varphi}=\normalbaselineskip=18pt\pmatrix { {\partial
  r\over\partial\varphi_2}\cos \theta-r\sin \theta{\partial
  \widetilde\theta\over\partial\varphi_2}&-r\sin\theta\cr
{\partial r\over\partial\varphi_2}\sin \theta+r\cos \theta
  {\partial\widetilde\theta\over\partial\varphi_2}&r\cos\theta\cr}, \quad B={\partial P\over\partial
  \varphi}=\normalbaselineskip=18pt\pmatrix { {\partial
  \rho\over\partial\varphi_2}\cos \psi&0\cr{\partial \rho\over\partial\varphi_2}\sin \psi&0\cr}$$
From (4.1) and the relation ${d\over dt}=\omega_1{\partial\over\partial\varphi_1}
+\omega_2{\partial\over\partial\varphi_2}$ 
on the torus $\Lambda$, we recover the fact that the focal points are
given by ${\partial r\over\partial\varphi_2}=0$, i.e. $\rho=0$. Moreover, the condition 
that $(C,B)$ is everywhere of rank 2 (see Leemma a.1~) yields
${\partial\rho\over\partial\varphi_2}\neq 0$ when $\rho=0$. The caustic $r=r_0$ consists here only in points of ${\cal F}_0$,
and the degeneracy is due to the radial symmetry of the metric. A particular case, with zero angular momentum,
was studied in [O-de-AlHa]. 
\medskip
\noindent {\it Remark}: It is known that due to the positive homogeneity of ${\cal H}$
with respect to the action variables, Euler formula gives $\omega_1I_1+\omega_2I_2={\cal E}$ (see [Ze]). Let us give another
simple proof of this. Considering the Lagrangian formulation associated with ${\cal H}$, we see that the energy of the (unit mass) 
particle moving along the geodesic flow is 
$${\cal E}=-\!\!\!\!\!\!\int\langle\dot x,p\rangle dt$$
where $-\!\!\!\!\!\!\int$ denotes $\lim_{T\to\infty}{1\over T}\int_0^T$. 
Substituting $\dot x=\langle \omega,\partial_\varphi\rangle x$,  we get
${\cal E}=\Sum_j\omega_j-\!\!\!\!\!\!\int\langle{\partial x\over\partial\varphi_j},p\rangle dt$.
Now if the frequencies are rationnally independent we can replace the time average by the average over angles variables,
so that by the definition of $I_j$~:
$${\cal E}=\Sum_j\omega_j (2\pi)^{-d}\int_{{\bf T}^d}\langle{\partial x\over\partial\varphi_j}x,p\rangle d\varphi=\Sum_j\omega_jI_j$$
\smallskip
Consider now Maupertuis-Jacobi correspondance allowing to switch from ${\cal H}$ to the hamiltonian $H=p^2-{1\over g(r)}$.
As a semiclassical example, we consider the Schr\"odinger operator with
potential $-u(r)$~:
$$ H=-{h^2\over r}{\partial\over \partial r}r{\partial\over \partial r}-{h^2\over r^2}
{\partial^2\over\partial\theta^2}-u(r)$$
We separate variables expanding as a  Fourier series in $\theta$, and rescale by $\sqrt r$, 
which leads to a family of
1-d semi-classical channel operators of the form $H_n=-h^2{\partial^2\over \partial r^2}+V_n(r)$ 
with effective potential $V_n(r)=h^2{n^2+1/4\over r^2}-u(r)$. 
Thus the angular momenta are given by $\psi=\psi_n=h(n^2+1/4)^{1/2}$, and the turning surfaces $r=r_0$ 
at energy $E$ by
$r^2(E+u(r))=\psi_0^2$. So because of the centrifugal barrier, the energy shell for $H_n$ 
can have a bounded component $0<r_1\leq  r\leq  r_2$ in configuration space, and another unbounded component 
$r\geq r_3$ (consider e.g.  $u(r)= 1/r+r>0$ and $E=0$.~) 
A torus $\Lambda$ corresponding to the angular momentum $\psi_n$ in the compact component of the energy surface,
carries a quasi-mode for suitable $E=E(h)$, and $E(h)$ turns in general into a resonance for 
hamiltonian $H$, whose imaginary part has to be computed.

A famous example is given by the dynamics in ${\bf R}^3$ of an electron in a Coulomb potential $-1/r$ , 
and interacting with a small, constant magnetic field of strenght $B$ (the constructions above should be extended to the 3-D case).
In suitable units, and cylindrical coordinates $(\rho,\theta,z)$, with dual variables $(p_\rho,p_\theta,p_z)$ the 
Hamiltonian takes the form~:
$$H=\bigl(p_\rho^2+p_z^2+{p_\theta^2\over\rho^2}\bigr)+B\rho^2-(\rho^2+z^2)^{-1/2}$$
The Coulomb singularity can be removed by introducing Jacobi or ``parabolic'' coordinates, see e.g. [Ts].
The Rayleigh-Schr\"odinger series for resonances  
near $E=1/4$ is Borel summable and can be computed perturbatively around $B=0$ (Zeeman effect). 
This (non semiclassical) problem has been completely solved in [HeSj], and lies far beyond the scope of our analysis.
Nevertheless, we would like to point out that resonances arise from connecting real tori 
(one of them is actually a cylinder, and lies over an annulus of the form $\Omega=\{r_3<\rho<+\infty\}$,~) 
through tunnel cycles in the classically forbidden region, see [DoSh] for a related problem. 
\medskip
\noindent {\bf b) The Kerr-Newman space time}.
\smallskip
We give another example of a metric for which we can find  some invariant (degenerate) tori and reparametrize
the natural phase space variables with action-angle variables by the procedure above.  
Namely, consider the geodesic flow associated with a simplified Kerr-Newman metric (see e.g. [Ch], or [H\"a] and references therein,~)
for which the black-hole angular momentum per unit mass is $a=0$ (Reissner-Nordstr\"om space-time),
$M$ its  mass and $Q$ its charge.
The Hamiltonian reads~:
$${\cal H}={1\over 2r^2}\bigl({r^4\tau^2\over\Delta}-{\varphi^{*2}\over\sin^2\theta}-\Delta\rho^2-\theta^{*2}\bigr)$$
The variables $(t,\tau,r,\rho,\theta,\theta^*,\varphi,\varphi^*)\in T^*{\bf R}^{1+3}$
are denoted collectively by $(P,X)$, and $\Delta=\Delta(r)=r^2-2Mr+Q^2$, 
is a second order polynomial with constant coefficients 
having 2 real and positive roots $r_1<r_2$. 
The region $r_1\leq r\leq r_2$ is called Boyer-Lindquist block $B_{II}$.
The equations of motion are of the form
$$\eqalign{
&\dot t={r^2\over\Delta}\tau, \dot \tau=0,\quad  \dot r=-{\Delta\over r^2}\rho, \quad \dot\rho=-{\partial{\cal H}\over\partial r}\cr
&\dot\theta=-{\theta^*\over r^2}, \quad  \dot\theta^*={1\over 2r^2}\partial_\theta{\varphi^{*2}\over\sin^2\theta}, \quad
\dot\varphi=-{1\over r^2\sin^2\theta}\varphi^*, \quad \dot \varphi^*=0\cr
}\leqno(4.7)$$
It is known that there are 4 integrals of motion, given by
$$p=2{\cal H}, E=\tau, L=-\varphi^*, {\cal K}=\theta^{*2}+{L^2\over\sin^2\theta}={r^4\tau^2\over\Delta}-\Delta\rho^2-pr^2$$
The energy surface is of course non bounded, but has some bounded connected components in the space-like region
$t=t_0, \tau=0$, we can call ``unphysical''.
The moment map restricted to $\tau=0$ is of the form $m=\bigl(p,L,{\cal K})$.
Let us find the actions variables and the fundamental frequencies $\omega=(\omega_1,\omega_2,\omega_3)$
on the corresponding torus $\Lambda$. We have $\langle P,dX\rangle=\tau dt+\rho dr+\theta^*d\theta+\varphi^*d\varphi$.

As a first fundamental cycle $\gamma_1$ in
$\Lambda$, we take $r=r_0\in\{r_1,r_2\}$), $\varphi^*=-L$. Integrating
(4.7) along $\gamma_1$, we find that $d\varphi$ and $d\theta$ are related by $d\varphi={\varphi^*\over\theta^*\sin^2\theta}d\theta$
so the first action variable takes the form~:
$$I_1={1\over 2\pi}\int_{\gamma_1}\langle P,dX\rangle={{\cal K}\over 2\pi}\int_0^{2\pi}
\bigl({\cal K}-{L^2\over\sin^2\theta}\bigr)^{-1/2}d\theta$$
which makes sense only if $L=0$. So the dimension of $\Lambda$ is 2, and $I_1=\sqrt{\cal K}$.

Let $\gamma_2$ the second  fundamental cycle be given by $\theta=\theta_0$, $r_1\leq r\leq r_2$,  this yields
another Clairaut type integral
$$I_2={1\over 2\pi}\int_{\gamma_2}\langle P,dX\rangle={1\over 2\pi}\int_{r_1}^{r_2}\rho dr=
{1\over 2\pi}\int_{r_1}^{r_2}\bigl({{\cal K}+pr^2\over -\Delta}\bigr)^{1/2}dr$$
where we have made use of $E=0$.

The last components of the degenerate (3+1)-torus in $t=t_0$ is then $I_3={1\over 2\pi}\int_0^{2\pi}\varphi^*d\varphi=0$,
and $\omega_3$ is undefined.
It is then clear from the Hamilton equations (4.7) that the description above of the motion for the geodesic flow with a
radial metric extends to this case without change.  In particular, formulae corresponding to (4.4) and (4.5) hold.
We shall not investigate further the case of degenerate tori (see [BeDoMa] for a general discussion.) 
Other ``solvable models'' are considered in [DuFoNo,Chap.7].
\medskip
\noindent {\bf c) Liouville metrics}.
\smallskip
These metric have been extensively studied ([Fo], [BNe], [Se], [Mat], \dots) for their nice property of being square integrable,
in the sense that the geodesic flow has another quadratic integral, i.e. a quadratic polynomial in $p$.  

We call {\it Liouville metric} on a domain $\Omega\subset{\bf R}^2$ a conformal metric given in $T^*\Omega$
by the hamiltonian ${\cal H}(x,p)=g(x)p^2$ with a quadratic (in momentum) second integral. 

We call {\it model Liouville metric} on $\Omega$ the Liouville metric where
$$g(x)=\bigl(u(x_1)+v(x_2)\bigr)^{-1}\leqno(4.8)$$
It is straightforward to check that, besides ${\cal H}(x,p)=E$, an additional integral of the motion is given by 
$$f(p,x)={v(x_2)p_1^2-u(x_1)p_2^2\over u(x_1)+v(x_2)}=F\leqno(4.9)$$ 

A classical result of Birkhoff (see [Fo,p.291]) says that in certain appropriate coordinates any 2-D Riemannian metric
whose geodesic flow has a additional integral quadratic in momenta is (locally) a model Liouville metric.

An interesting case is when $g$ is globally of the form (4.8) on $\Omega$. Thus the hamiltonian system decouples
completely into two 1-d hamiltonians with a potential
$p_1^2=u(x_1)E+F, p_2^2=v(x_2)E-F$, and moment map $m=(E,F)$.
The corresponding semiclassical Schr\"odinger
operator is also completely integrable, in the sense that the spectral problem reduces in finding the joint
spectrum of 2 commuting 1-D Schr\"odinger operators, see e.g. [KMS] for a study of the spectra
of Laplace-Beltrami operators on a Liouville surface, and also 
[DoSh] for tunneling effects. 

Moreover, square integrable conformal metrics when $\Omega$ is an annulus, are globally,
conformally equivalent to model Liouville metrics in the following sense~: 
\medskip
\noindent {\bf Proposition 4.1} [Mat]: Let ${\cal H}(x,p)=g(x)p^2$ be a smooth conformal square integrable metric on
a compact domain $\Omega$ of the $x$-plane diffeomorphic to an annulus, and 
$\Lambda\subset T^*\Omega$ be an integral
lagrangian manifold for the hamiltonian flow such that $\pi:\Lambda\to \Omega$ is surjective ($\Lambda$ has 
no ``holes'', i.e. there is no forbidden region above $\Omega$.~)
Then there exists a bijective conformal map $\alpha:\Omega\to \widetilde \Omega$ (depending on $\Lambda$) where 
$\widetilde \Omega$ is an annulus of the $x$-plane, and
such that the image of ${\cal H}$ through $\alpha$ becomes a model Liouville metric.
\smallskip
One can also compute action integrals $I$ and corresponding frequencies $\omega$ as in Part a). In particular we have the
relation $\omega_1I_1+\omega_2I_2=2E$.
At the level of operators, we use Weyl $h$-quantization of the form ${\cal H}(x,hD)=hD_xg(x)hD_x$, so that the
diffeomorphism $\alpha$ takes ${\cal H}(x,hD)$, denoting by $y$ the new space coordinates, to 
$\widetilde{\cal H}(y,hD)=hD|\nabla\alpha|^2g\circ\alpha(y)hD$. This operator is of special interest in the theory of surface
water waves captured by coasts and bottom relief irregularities.

Consider now the following semiclassical ``Maupertuis problem''~: let 
${\cal H}(x,p)=g(x)p^2$ and $\Lambda\subset T^*\Omega$ be as in Proposition 4.1, with ${\cal H}|_\Lambda=1$. 
Let $H=-h^2\Delta-1/g(x)$, and $A=A(x)$ be the unitary Fourier Integral Operator (FIO) on $L^2(\Omega)$ implementing 
$\alpha$ (suitably extended to the whole space). Then $\widetilde H=A^*HA$ is a self-adjoint $h$-PDO 
on $L^2(\widetilde \Omega,|\nabla\alpha|^2dy)$
with principal symbol
$\widetilde H_0(y,p)= p^2-1/\widetilde g(y)$
(here $\widetilde g=g\circ\alpha$ is the model Liouville metric
on $\widetilde \Omega$ given in Proposition 4.1.~) There is no sub-principal symbol. 
Moreover, $\Lambda$ is an invariant lagrangian manifold for $X_{\widetilde H}$. So we may apply Theorem 3.2
to construct quasi-modes for $\widetilde H$. 
In the following subsection we shall introduce another Maupertuis correspondance for this geodesic flow.
\medskip
\noindent{\bf d) The shallow water waves}.
\smallskip 
Let $\Omega\subset{\bf R}^2$ (the ``basin'') be the complement of a bounded set (the ``island''), so that $\Omega$
is diffeomorphic to an (unbounded) annulus.
In the shallow water waves theory [DoZh] one constructs a $h$-PDO with symbol $L(p,x;h)=L_0(p,x)+hL_1(p,x)+\cdots$. 
The principal symbol of $L$ (dispersion relation) on $T^*\Omega$ reads~:
$$L_0(p,x)=|p|(1+\mu p^2)\tanh(D(x)|p|)\leqno(4.10)$$ 
where $D$ is a smooth positive function, describing the depth of the basin, and $\mu>0$ a small parameter.
Quasi-modes of $L$ represent the  waves trapped by the island.
By a Maupertuis-Jacobi correspondance, we shall construct trapped modes for $L$ at energy $E$ from
lagrangian tori of Liouville metrics as above, by choosing the depth profile in an appropriate way. 
Since Liouville metrics are actually generic among metrics with a quadratic second integral 
[BeN], this result could be of practical interest.

For simplicity we shall assume $\mu=0$, but it is not hard to see that our constructions may be carried out perturbatively, due to
implicit functions theorem, for $\mu>0$ small enough.
Note that, if $D(x)$ is positively homogeneous of degree 1, then integrability of $L_0(p,x)=|p|\tanh(D(x)|p|)$ 
at energy 1 implies integrability
at all energies $E>0$ (make the symplectic change of coordinates $(x',p')=(Ex,E^{-1}p)$.~)
In general case however, $L_0(x,p)$ has no reason for being integrable at any energy $E$, but finding suitable profiles
that give raise to invariant tori can be achieved by the following Maupertuis correspondance.

Namely, multiply relation $L_0(p,x)=E$  (for $\mu=0$) by $D(x)$, 
and set $z=|p|D(x)$, $\phi(z^2)=z\tanh z$. We denote by $\psi:{\bf R}_+\to{\bf R}_+$ the
inverse of $\phi(z^2)$, which is a smooth increasing function, so that (4.10) can be rewritten as 
$${\cal H}(x,p)=g(x,E)p^2=1, \quad g(x,E)={D^2(x)\over \psi(ED(x))}\leqno(4.11)$$
We have the elementary~:
\medskip
\noindent{\bf Lemma 4.2}: On the energy surface ${\cal H}(x,p)=1$, the Hamilton vector fields $X_{L_0}$ and $X_{\cal H}$ are
related by $X_{\cal H}={\psi(ED(x))\over D(x)\psi'(ED(x))}X_H$, 
and this induces a reparametrization of the integral curves of
$X_H(X(t),P(t))$ given by Maupertuis-Jacobi correspondance
$$dt={\psi(ED(x))\over D(x)\psi'(ED(x))}d\tau\leqno(4.12)$$
\smallskip 
So we need look for depth profiles $D(x)$ (depending also on $E$,~) for which $g(x,E)p^2$ is integrable. 
Clearly the function $G:{\bf R}_+\to[0,1[$ defined by 
$G(D)={D^2\over \psi(D)}$ is a smooth increasing function, satisfying $G(D)=D(1-{1\over 3}D+\cdots)$ as $D\to 0$, and 
$\lim_{D\to\infty}G(D)=1$,
so the depth profile $D(x)=D(x,E)$ can be chosen as the solution of 
$g(x,E)=g(x)=E^{-2}G(ED(x,E))$, where $g(x)$ is one of the square integrable metrics given in Proposition 4.1. 

Now if the geodesic flow ${\cal H}(x,p)=g(x,E)p^2$ is integrable and admits invariant tori, the procedure
set up in Sect.3 can be extended to construct quasi-modes for 
Weyl $h$-quantization $L^w(x,hD)$ of hamiltonian $L(x,p;h)$, for which eigenfunctions, or quasi-modes,
describe again trapped waves near energy $E$. Note however that $L$ is a $h$-PDO of more general type 
than a Schr\"odinger operator, but all our constructions, and in particular the results of Appendix,
apply to this setting, as we pointed out after Theorem 3.2. Thus we can state our main result as~:
\medskip
\noindent{\bf Theorem 4.3}: Consider the $h$-PDO with symbol $L(p,x;h)=L_0(p,x)+hL_1(p,x)+\cdots$ with $L_0$ as in (4.10)
defined on an open set $\Omega\subset{\bf R}^2$ diffeomorphic to an annulus. For all $E>0$, and all smooth Liouville metric 
${\cal H}(x,p)=g(x,E)p^2$ on $T^*\Omega$, there exists a ``depth profile'' $D(x)=D(x,E)$ on $\Omega$, such that the pair of
Hamiltonians $(L_0,{\cal H})$ have the same trajectories at energies $(E,1)$, both parametrizations being related by (4.12).
In particular, ${\cal H}(x,p)$ being integrable, $(L_0,{\cal H})$ have a common family of lagrangian tori $\Lambda$ 
for energies $(E,1)$, and Theorem 3.2 allows to construct a sequence of quasi-modes for $L(x,hD_x;h)$ near 
energy $E$. 

\bigskip
\centerline{\bf Appendix :  Regular WKB constructions and Maslov theory}
\medskip
In this Section we recall some well known facts about semi-classical quantization by standard Maslov theory~; 
we refer to [Ar1,2], [So], [CdV1], [H\"o1,2], [Laz], [Le], [Li], [Du1,2], [GuSt], [Iv], [We], [MeSj], 
[BaWe], [DoZh], [deG]\dots for more details.  In Part A we give a rather superficial account for singularities of a 
lagrangian manifold diffeomorphic to a torus, and paramerized globally by angle variables 
$\varphi\in{\bf T}^d$ (essentially when $d=2$)~; 
then we derive a simple formula for Maslov index. In Part B we recall some known facts about WKB theory, 
when the tangent space to the lagrangian manifold contains $X_H$,
as a preliminary step to Sect.3 for the perturbative case.
In Part C we combine these results to get Bohr-Sommerefeld-Maslov quantization condition. 
\smallskip
\noindent{\bf A. Maslov index and singularities for a lagrangian immersion of the torus}.
\smallskip
Let $M$ be a smooth Riemannian manifold,
(we shall think essentially of $M={\bf R}^d$ or the flat torus
$M={\bf T}^d$ in the case of the geodesic flow on a Liouville surface,~)
and $\iota:{\bf T}^d\to T^*M$ a smooth lagrangian immersion,
we parametrize $\Lambda=\iota({\bf T}^2)$ by $p=P(\varphi), x=X(\varphi)$. To comply with usual notations,
we identify already ${\bf T}^d$ with its image through $\iota$, so that the Lagrangian
immersion reads now $\iota:\Lambda\to T^*M$.

We begin with a short description of lagrangian immersions. Contrary to local lagrangian singularities,
the global geometry of lagrangian tori, especially 
those arising from integrable systems is not so commonly discussed (see however [O-de-AlHa], [DoZh] and references therein.~) 
We call {\it focal point} a point $y\in\Lambda$ where
$\pi_*:T\Lambda\to TM$ is singular, and {\it caustics} the projection ${\cal C}$ of the set
of focal points onto $M$. 
We introduce the jacobian matrices $B={\partial P\over\partial\varphi}$ and $C={\partial X\over\partial\varphi}$, 
and recall the following standard result (see e.g. [DoZh]) 
\medskip
\noindent {\bf Lemma a.1}: With the notations above, we have~:

\noindent  1) the matrix $(B, C)$ is of rank $d$.

\noindent  2) the matrices ${}^tCB$, $BC^{-1}$ and $CB^{-1}$ are symmetric (whenever well defined, otherwise replace 
e.g. $B^{-1}$ by $\det (B) B^{-1}$.~)

\noindent  3) $C\pm iB$ is non degenerate.
\smallskip
\noindent {\it Proof}: The 2 first statements easily follow from the definition of a Lagrangian manifold, and the 3:rd one from computing
the hermitian scalar product $\bigl(\xi|({}^tBC-{}^tCB)\xi\bigr)$ when $\xi$ belongs to the kernel of $C\pm iB$.
$\clubsuit$

Taking variations, we see that tangent vectors $(\delta X,\delta P)\in
T\Lambda$ are given by $\delta P=B(y)\delta\varphi$, 
$\delta X=C(y)\delta\varphi$, so $\delta P=BC^{-1}(y)\delta X$ when $T_y\Lambda$ is 
transverse to the fiber, or $\delta X=CB^{-1}(y)\delta P$
when $T_y\Lambda$ is transverse to the zero section of $T^*M$. So $\phi''(x)=BC^{-1}(y)$, when $\Lambda$ is parametrized by the 1-form
$d\phi=\iota^*(pdx)$, while $\psi''(p)=CB^{-1}(y)$, when $\Lambda$ is parametrized by the 1-form
$d\psi=\iota^*(-xdp)$. 

For simplicity restrict now to the 2-d case, and let $y$ be a focal point. By property 1) of the Lemma, either
$B(y)$ is of maximal rank, or both $C(y)$ and $B(y)$ are of rank 1. 

In the first case, either $C(y)$ is of rank 1 (which, under some transversality condition, is the generic case,
and corresponds to the simple {\it fold} $A_2$,~)
and $\Lambda$ near $y$ intersects (tangentially) the fiber $VM$ (the vertical space of $T^*M$)
along a line, so ${\cal C}$ is of codimension 1,
or $\Lambda$ is tangent to the fiber of $T^*M$ at $y$, and we call $y$ an {\it umbilic}. The set of umbilics
denoted by ${\cal F}_1$ is generically discrete, and thus does not contribute to Maslov index.

In the second case there are $\lambda, \mu\in{\bf  R}
\setminus 0$ such that ${\partial X\over\partial\varphi_2}=\lambda
{\partial X\over\partial\varphi_1}$ and ${\partial P\over\partial\varphi_2}
=\mu{\partial P\over\partial\varphi_1}$ at $y$. Identifying the off-diagonal terms
of $^tBC$, which is symmetric by property 2) of the Lemma,
we find that either $\lambda=\mu$ or ${\partial X\over\partial\varphi_1}
\perp {\partial P\over\partial\varphi_1}$ (in the sense of the euclidean
structure on $T^*M$, since $M$ is flat.~) 
But $\lambda\neq\mu$ since otherwise the complex
matrices $C\pm iB$ would be degenerate, which violates
property 3) of the Lemma. So ${\partial X\over\partial\varphi_1}$
and ${\partial X\over\partial\varphi_2}$ are colinear, and orthogonal
to both ${\partial P\over\partial\varphi_1}$ and 
${\partial P\over\partial\varphi_2}$. 

We call $y$ a {\it cusp}. The corresponding lagrangian singularity is generically the Whitney crease $A_3$,
and we denote the set of cusps by ${\cal F}_0$.

Assume now $\Lambda\subset\Sigma_E$
where $\Sigma_E$ is a level set for an Hamiltonian of the form $H=p^2+V(x)$.
Differentiating, we get $2\langle P,{\partial P\over\partial\varphi_j}\rangle
+\langle\nabla V,{\partial X\over\partial\varphi_j}\rangle=0$  
on $\Lambda$ at $y$, $j=1,2$, and since $\lambda\neq\mu$, these relations imply
${\partial X\over\partial\varphi_j}\perp\nabla V(X)$ and 
$P\perp{\partial P\over\partial\varphi_j}$, $j=1,2$. So if both $C(y)$
and $B(y)$ are of rank 1, the vectors $P$,
${\partial X\over\partial\varphi_1}$ and 
${\partial X\over\partial\varphi_2}$ are colinear, and all orthogonal to
$\nabla V$, ${\partial P\over\partial\varphi_1}$ and
${\partial P\over\partial\varphi_2}$. 

Generically (i.e. without additionnal symmetries)
the set ${\cal F}_0$ of such focal points $y$ is discrete, or even empty, and like umbilics, do not
contribute to Maslov index. The same argument applies if $H=p^2g(x)$ (geodesic flow) or 
$H=(p_1-x_2)^2+(p_2+x_1)^2+V(x)$. See however Sect.4 for the case of rotational symmetry. 
\medskip
\noindent {\it Example a.1}: The harmonic oscillator $H=p^2+V_0(x)$, 
$V_0(x)=\omega_1^2x_1^2+\omega_2^2x_2^2$. Although standard action-angle variables are singular at energy $E=0$, 
they are often best suited for describing the singularities at caustics.
For all $E>0$, there are 4 (hyperbolic) umbilics, located at the vertices of the caustics (a rectangle).
For small energies $E>0$, these umbilics are stable under perturbations of the form $V(x)=V_0(x)+{\cal O}(|x|^3)$.  
\smallskip
\noindent {\it Example a.2}: The magnetic hamiltonian in ${\bf R}^2$ of the form 
$H_A=(p-A(x))^2+V(x)$ (written in the minimal coupling setting) with $V$ as above, and  
$A(x)={\cal O}(|x|^2)$ near $x=0$. Then umbilics (when $A(x)\equiv 0$) metamorphose into cusps, and singularities
are precisely Whitney creases, see [KaRo]. For more general magnetic hamiltonians in $T^*M$, $M={\bf R}^3$, it is convenient to 
change the minimal coupling formulation, and introduce the non-degenerate (magnetic) 2-form $\sigma_B=\sigma-B(x)$ with
$B(x)=dA(x)$, which makes of $T^*M$ a new symplectic manifold. 
When $V(x)=\Sum_jV_j(x_j)$, for smooth potentials $V_j(x_j)$, it is easy 
to see that $I_j=(p_j-A_j(x))^2+V_j(x_j)$ Poisson commute for $\sigma_B$, and thus can be chosen as action-variables.
\smallskip
\noindent {\it Example a.3}: A radially symmetric metric ${\cal H}=g(|x|)p^2$ on an annulus $0<r_1\leq r\leq r_2$.
In general, the whole caustics lies in ${\cal F}_0$ (see again Sect.4).
(This emphasizes the peculiarity of the harmonic oscillator among integrable systems.~)
\smallskip
\noindent {\it Example a.4}: The geodesic flow on an ellipsoid is conformally equivalent, in the upper-half space, to 
a Liouville metric ${\cal H}=g(x)p^2$ on a rectangle $R=]0,T_1[\times]0,T_2[$.  
There are 4 umbilics going to the
vertices of $R$, and the induced metric near an umbilic takes the form, in local coordinates
$1/g(x)= G(x_1^2)-G(-x_2^2)=(x_1^2+x_2^2)A(x_1^2-x_2^2,2x_1x_2)$, where $A(0,0)>0$ (see [CdVVu,Sect.3]
and references therein.~) 
\medskip
Now we can pass to Maslov theory and consider first the linear situation. 
Recall the real phase space $T^*{\bf R}^d$, in addition with its symplectic
and euclidean structures, is endowed with the complex structure ${\cal J}(p,x)=
(-x,p)$,~; we identify $T^*{\bf R}^d$ with ${\bf C}^d$ and its zero section
${\bf R}^d_p\oplus 0$ with the totally real subspace ${\bf R}^d_p$. 

Let ${\cal L}(2d)$ be the lagrangian grassmannian of $T^*{\bf R}^d$,
we know that $U(d)$ (the unitary group on ${\bf C}^d$) preserves all these structures and
acts transitively on ${\cal L}(2d)$. We can 
identify ${\cal L}(2d)$ with $U(d)/O(d)$, $O(d)$ (the orthogonal group on ${\bf R}^d$) being the stabilizer
of a given ${\bf R}^d\approx\Lambda\in{\cal L}(2d)$ under the action of $U(d)$. 
Equivalently, if $\Lambda={\cal U}_1({\bf R}^d)={\cal U}_2({\bf R}^d)$
(with the identification above of the zero section,~) ${\cal U}_j\in U(d)$, then 
${\cal U}_2={\cal U}_1{\cal O}$, ${\cal O}\in O(d)$. Since $\det ^2{\cal O}=1$,
$\det ^2{\cal U}_1=\det ^2{\cal U}_2$ only depends on $\Lambda$. The map
$\det ^2:U(d)\to{\bf S}^1$ induces a fibration ${\cal L}(2d)\to{\bf S}^1$,
with 1-connected fiber $SU(d)/SO(d)$, giving an isomorphism of fundamental
groups~: $\pi_1({\cal L}(2d))\approx \pi_1({\bf S}^1)\approx {\bf Z}$. 
Passing to homology and dualizing, we obtain a natural homomorphism~:
$H^1({\bf S}^1;{\bf Z})\to H^1({\cal L}(2d);{\bf Z})$.
The image $\mu=\mu_d$ of the canonical generator of $H^1({\bf S}^1;{\bf Z})$ under
this map is called the universal Maslov class. Explicitely, along the path 
${\bf S}^1\to{\cal L}(2d), t\mapsto\Lambda_t={\cal U}_t({\bf R}^d)$,
${\cal U}_t\in U(d)$, we have
$$d\mu(\Lambda_t)=d\mu_t={1\over 2\pi}d(\arg \det {\cal U}_t^2)
\leqno(a.3)$$
Consider now the symplectic vector bundle $\pi:T^*M\to M$ over a manifold $M$,
with a lagrangian subbundle $\Lambda\to T^*M$.
In general, the automorphism group of $T^*M$ does not act transitively on the
lagrangian subbundles, but a pair of transverse lagrangian subbundles  
can be related as follows. Let $\iota:\Lambda\to T^*M$ be a lagrangian  
immersion,  then the symplectic vector bundle $\iota^*T(T^*M)$ over
$\Lambda$ has 2 lagrangian subbundles $L=\iota_*T\Lambda$ and $L_2=\iota^*VM$
(the ``vertical'' subbundle). The same consideration holds when replacing
$L_2$ by the zero-section $L_1$ of $T^*M$. Maslov class $\alpha=\mu_{\Lambda,\iota}$ of $(\Lambda,
\iota)$ is defined as $\mu_{\iota,\Lambda}=\mu(L,L_2)\in 
H^1(\Lambda;{\bf Z})$, which is the pull-back of the universal Maslov class
$\mu$  by Gauss map $G:\Lambda\to{\cal L}(2d)$,
$y\mapsto G(y)=\iota_*T_y\Lambda$. 

For $y\in\Lambda$, let ${\cal U}(y)$ be the unitary transform generating 
the lagrangian plane $T_y\Lambda$. Possibly after changing $d\mu_t$ into $-d\mu_t$ in (a.3) it is clear that, at least for $d=1$, 
${\cal U}^2(y)$ is given globally by Cayley transformation 
$${\cal U}^2(y)=(C+iB)(C-iB)^{-1}(y), \quad d=1 \leqno(a.4)$$
\medskip
\noindent {\it Example a.5}: In case $\Lambda$ is the product of $d$ circles 
$p_j^2+x_j^2=E_j$ parametrized by $(p_j,x_j)=(\sqrt {E_j}\sin\varphi_j,\sqrt {E_j}\cos\varphi_j)$,
(a.3) and (a.4) give by integration Maslov index $\alpha=\mu(L,L_2)=(2,\cdots,2)$. 
\smallskip
We still show below how (a.4) holds ``at the infinitesimal level'' and allows to compute 
the jump of Maslov index $\mu(L,L_2)\in H^1(\Lambda,{\bf R})$ passing a caustics.
See also [Ma2], [DoZh] for another and more explicit derivation of $\mu(L,L_2)$. 

Moreover, from the discussion after Lemma a.1, the RHS of (a.4) equals
$(I+iBC^{-1})(I-iBC^{-1})^{-1}(y)$ when $\Lambda$ is transverse at $y$ to $L_2$, and 
$-(I-iCB^{-1})(I+iCB^{-1})^{-1}(y)$ when $\Lambda$ is
transverse at $y$ to $L_1$. 
Thus, because  $BC^{-1}(y)$ and $CB^{-1}(y)$ are symmetric, 
except perhaps on an exceptional set which we identified with
${\cal F}_0$ in the 2-d case, the RHS of (a.4) is well defined as a unitary matrix acting on lagrangian planes, i.e.
the fibers of $T\Lambda$. So we are in the situation of [H\"o2,p.156], up to the sign in front
of $-(I-iCB^{-1})(I+iCB^{-1})^{-1}(y)$, which we may ignore if $d$ is even (since only $\det {\cal U}^2(y)$ matters), or if
we forget about orientation. So we write
${\cal U}^2(y)=(I+iBC^{-1})(I-iBC^{-1})^{-1}(y)$ (resp. ${\cal U}^2(y)=(I-iCB^{-1})(I+iCB^{-1})^{-1}(y)$. 

Following [So] (see also [GuSt]) we shall identify these unitary matrices with the lagrangian planes on which they act,
and introduce also the covering space $\widetilde {T\Lambda}$ of $T\Lambda$ by adding a new (angular) variable $\theta$ 
which stands for $\arg \det {\cal U}$. So in $\widetilde T\Lambda$ we have the freedom
to consider a path of ``fictitious'' lagrangean planes $(\widetilde L_t)_{0\leq t\leq 1}$ which are not tangent to $\Lambda$, 
except for $t=0,1$.
A similar argument, consisting in interpolating between $C+i\e B$ and $C-i\e B$, was used in [DoZh]. 
Now we argue as in [H\"o2], and 
to start with, restrict to the case where ${\cal C}_{y_0}$ has codimension 1 in $\Lambda$ and $B(y)$ is of maximal rank $d$ 
near $y_0$. 

Consider first a path $\bigl(\gamma_1(t),\dot\gamma_1(t)\bigr)\in T\Lambda$, $-\epsilon_0\leq t\leq \epsilon_0$, with
$\dot\gamma_1(t)\in T_{\gamma_1(t)}\Lambda=\tau(t)$, and such that $\dot\gamma_1(0)$ belongs to the fiber. 
So in the terminology of [GuSt,p.120], $\tau(-\epsilon_0)$ and $\tau(\epsilon_0)$, 
lie in different components of the projective lines ${\cal L}_{L_2}$
(corresponding to the fiber $L_2$) and ${\cal L}_{L_2}$ (corresponding to the zero-section $L_1$). We lift the path 
$(\gamma_1,\dot\gamma_1)\bigr)$ in $\widetilde {T\Lambda}$ as $(\widetilde\gamma_1,\widetilde{\dot\gamma}_1)$. 
Then along this path, $B(y)|_{y=\gamma_1(t)}$ is always invertible, and 
${\cal U}^2(y)=(I-iCB^{-1})(I+iCB^{-1})^{-1}(y)$. The square of the determinant of ${\cal U}_t$ is
$\prod _j(1-i\beta_j(t))(1+i\beta_j(t))^{-1}$ where $\beta_j(t)$ are the real
eigenvalues of $CB^{-1}$. Since $\arg (1-i\beta_j(t))=
-\arg (1+i\beta_j(t))$ (we choose $\arg z$ so that it vanishes for $z>0$) we get
$\int_{\widetilde\gamma_1} d\mu_t={1\over \pi}\Sum_j\arg (1-i\beta_j(\epsilon_0))-\arg (1-i\beta_j(-\epsilon_0))$.

Choose next a path $(\widetilde\gamma_2(t),\widetilde{\dot\gamma}_2(t))\in\widetilde {T\Lambda}$, 
with $\theta$ between $\arg \det {\cal U}(\gamma(\epsilon_0))$ and arg det ${\cal U}(\gamma(-\epsilon_0))$, and
whose projection starts at $\tau(\epsilon_0)$ 
and ends at $\tau(-\epsilon_0)$. We use
instead ${\cal U}_t^2(y)=(I+iBC^{-1})(I-iBC^{-1})^{-1}(y)$. 
The square of the determinant of ${\cal U}_t$ is now
$\prod_j(1+i/\beta_j(t))(1-i/\beta_j(t))^{-1}$, and 
$\int_{\widetilde\gamma_2}\mu_t={1\over \pi}\Sum_j\arg (1+i/\beta_j(1))-
\arg (1-i/\beta_j(0))$. Since for real $\beta\neq 0$, $\arg (1-i\beta)-
\arg (1+i/\beta))=-\pi/2$ if $\beta>0$, $=\pi/2$ if $\beta<0$, we find~:
$$\oint d\mu_t=\bigl(\int_{\widetilde\gamma_1}+\int_{\widetilde\gamma_2}\bigr)d\mu_t=
{1\over 2}\bigl(\sgn  (BC^{-1})(\epsilon_0)-\sgn  (BC^{-1})(-\epsilon_0)\bigr)$$
Here $\oint$ denotes the integral over the loop $\widetilde\gamma=\widetilde\gamma_1\vee\widetilde\gamma_2$ in $\widetilde {T\Lambda}$, 
and we have used that for $t\neq 0$, $BC^{-1}$ and $CB^{-1}$ are well defined and $\sgn BC^{-1}=-\sgn CB^{-1}$. 
This gives the required formula
$$\langle\widetilde\gamma,\alpha\rangle={1\over 2}\bigl(\sgn  (BC^{-1})(\epsilon_0)-\sgn  (BC^{-1})(-\epsilon_0)\bigr)\leqno(a.6)$$
When ${\cal F}_0$ is a discrete set it doesn't  contribute to Maslov index,
so formula (a.6) gives generically the jump of Maslov index  passing a caustic. But in some cases as the radial symmetric
potentials in $d=2$ considered in Sect.4, there are caustic sets where both $B(y)$ and $C(y)$ have rank 1, and 
the constructions above fail.  
The argument of [DoZh] covers most of these exceptional cases, but we can still adapt this one as follows.

Actually, since we can construct a subbundle transversal to both $L_2=\iota^*VM$ and $L=\iota_*T\Lambda$,
we know that near any $y_0=(p_0,x_0)\in\Lambda$, there is a local diffeomorphism $\chi$ (in the variable $x$ alone), such that
$\Lambda$ is parametrized by a non-degenerate phase function of the type $S(p,x)=\langle p,x\rangle+H(p)$, i.e.
$\Lambda=\{(p,x) \ : d_pS(p,x)=0, p\in\neigh (p_0)\}$. Moreover, $H''(p)$ is symmetric and invertible.
\medskip
\noindent {\it Example a.6}: Let $\Lambda\subset T^*{\bf R}$ be the circle $y^2+\eta^2=1$, after the symplectic change of 
coordinates $y=\chi(x)=e^x-1, \eta={}^t(\chi'(x))^{-1}\xi=e^{-x}\xi$ near $(y,\eta)=(0,1)$, $\Lambda$ rewrites as
$x=\xi-1+{5\over 6}(\xi-1)^3+{\cal O}((\xi-1)^4)$. 
\smallskip
When $\Lambda$ is of the form $(P(\varphi),X(\varphi))$, $\varphi\in{\bf T}^2$, one can locally use $p$ as local coordinates,
and make a symplectic transformation of the form $(p,x)\mapsto(p+\bigl((H''(p)\bigr)^{-1}x, x)$. 
The tangent space to $\Lambda$ is of the form $(\delta p-\bigl(H''(p)\bigr)^{-1}\delta x, \delta x)$
so we are reduced  to replace $(C,B)$ above by $(C,B-\bigl(H''(p)\bigr)^{-1}C)|_{p=P(\varphi)}$, and so
$BC^{-1}$ by $BC^{-1}-(H'')^{-1}$. By (a.6)
$$\langle\widetilde\gamma,\alpha\rangle={1\over 2}\bigl(
\sgn (BC^{-1}-(H'')^{-1})(\epsilon_0)-\sgn  (BC^{-1}-(H'')^{-1})(-\epsilon_0)\bigr)$$
and as in [H\"o2] this formula rewrites
$$\langle\widetilde\gamma,\alpha\rangle={1\over 2}\bigl(
\sgn \pmatrix{-H''(-\epsilon_0)&\Id \cr \Id 
&-BC^{-1}(-\epsilon_0)}-\sgn \pmatrix{-H''(\epsilon_0)&\Id \cr \Id &-BC^{-1}(\epsilon_0)}\bigr)
\leqno(a.7)$$
This formula remains valid by continuity even if $H''$ fails to be invertible, provided the matrices are non singular,
i.e. so long $T_{y(\pm\epsilon_0)}\Lambda$ is tranverse to the fiber and zero-section of $T^*M$.

\bigskip
\noindent {\bf B. The oscillating principal symbol and invariant densities}.
\smallskip
Here we review the construction of the principal symbol of asymptotic eigenfunctions of $H$, 
microlocalized on the invariant lagrangian manifold $\Lambda$, parametrized (globally) by $(p,x)=(P(\varphi),X(\varphi))$, 
$\varphi\in{\bf T}^d$. For a comprehensive treatment, in the framework
of Fourier Integral Operators calculus, we refer for instance the reader to the first Chapter of the book [Iv] by V.Ivrii which,
together with the Lecture Notes [BaWe] by S.Bates and A.Weinstein, we took as a first guide to 
this exposition. See also [DoZh].

So let $H(x,hD_x)$ be the semi-classical quantization of $H(p,x)$,
$\iota:\Lambda\to T^*M$ a lagrangian immersion in $H(p,x)=E$~; we look for 
an approximate solution mod ${\cal O}(h^2)$ of $H(x,hD_x)u=E(h)u$ with $E(h)=E+{\cal O}(h)$. 
For simplicity, we shall explicit the computations by assuming  $H(p,x)=p^2+V(x)$ so that 
$H(x,hD_x)=-h^2\Delta+V(x)$ is a Schr\"odinger operator, but everything carries over to more general $h$-PDO,
as in the case of linear water waves with dispersion relation given by (4.10). 
For the moment however, we shall assume that $H$ has no sub-principal symbol.

Near $x_0\notin{\cal C}$, we try a solution of simple WKB type, of the form $u(x,h)=a_0(x,E,h)$ $e^{i\phi(x)/h}$, where
$\phi$ solves Hamilton-Jacobi equation $H(x,\nabla\phi(x))=E$, i.e. $\phi(x)=\int_{x_0}^xpdx$, or $d\phi=\iota^*(pdx)$.
This can be done so long $\Lambda$ is projectable. To determine $a(x,h)=a_0(x,E)+ha_1(x,E)+\cdots$ we begin with
the first transport equation 
$$\Sum_j\phi'_j(x)\partial_{x_j}a_0(x,E)+{1\over 2}\Delta\phi(x)a_0(x,E)=0\leqno(b.2)$$
which expresses that $a_0^2\pi_{x*}(X_H)$ (the $x$-projection of the Hamiltonian vector field of $H$) is divergence-free
for the canonical density $|dx|$ on ${\bf R}^d$, but which we write more invariantly as 
$${\cal L}_{X_H|_{\Lambda}}(a_0(x,E)|dx|^{1/2})=0\leqno(b.3)$$
Here ${\cal L}_{X_H|_{\Lambda}}$ denotes the Lie derivative acting on half-densities. 
As $X_H$ is tangent to $\Lambda$, and the Lie derivative invariant under diffeomorphism, this equation is satisfied
iff the pull-back of $a_0(x,E)|dx|^{1/2}$ via the projection $\pi$ is invariant under the flow of $X_H$. 
When the hamiltonian flow is quasi-periodic on $\Lambda$, and $y=(p,x)=(P^0(\varphi),X^0(\varphi))$
it is readily seen that the solutions of (b.3) are of the form 
$$a_0(x,E)=\Const |\det {\partial X^0\over\partial\varphi}|^{-1/2}\leqno(b.4)$$
Next we analyze the situation near a focal point $y$, and begin with the other extreme, i.e. 
rank $d\pi_p(y)=d$. Near $y$, $\Lambda$ has a simple fold and the caustic is a line~;  if $d=2$, then
$y\notin{\cal F}_0$ where ${\cal F}_0$ was defined in (a.2). 
We make a $h$-Fourier transform, which takes formally $H(x,hD_x)=-h^2\Delta+V(x)$ to
$\widetilde H(p,-hD_p)=p^2+V(-hD_p)$. 
The action is now $\psi(p)=\int^p -xdp$, i.e. $d\psi=\iota^*(-xdp)$, and solves Hamilton-Jacobi equation 
$p^2+V(-\nabla\psi(p))=E$. So we look for $u(x,h)$ of the form 
$$I(b,\psi)(x,h)=(2\pi h)^{-d/2}\int e^{i(xp+\psi(p))/h}b(p,E,h)dp\leqno(b.5)$$
and formally $(H(x,hD_x)-E)u(x,h)=\int e^{i(xp+\psi(p))/h}(p^2+V(x))b(p,E,h)dp$. Using Ha- milton-Jacobi equation,
Taylor formula gives
$$p^2+V(x)-E=\Sum_j\lambda_j(x,p,E)(x_j+\partial_{p_j}\psi(p))\leqno(b.6)$$
where $\lambda_j(x,p,E)$ are smooth functions near
$x+\partial_p\psi(p)=0$, and because we work near a simple fold, we can assume, possibly after a rearrangement
of coordinates, that $\lambda_1|_{x+\partial_p\psi(p)=0}\neq 0$. We look for a symbol $b(p,E,h)=b_0(p,E)+hb_1(p,E)+\cdots$ 
such that there are
amplitudes $a^{(j)}(x,p,E,h)=a^{(j)}_0(x,p,E)+ha^{(j)}_1(x,p,E)+\cdots$ verifying
$$e^{i(xp+\psi(p))/h}(p^2+V(x)-E)b(p,E,h)=\Sum_jhD_{p_j}\bigl(e^{i(xp+\psi(p))/h}a^{(j)}(x,p,E,h)\bigr)\leqno(b.7)$$
At zeroth and first order in $h$, we find respectively
$$\eqalign{
&(p^2+V(x)-E)b_0(p,E)=\Sum_j(x_j+\partial_{p_j}\psi(p))a^{(j)}_0(x,p,E)\cr
&(p^2+V(x)-E)b_1(p,E)=\Sum_j(x_j+\partial_{p_j}\psi(p))a^{(j)}_1(x,p,E)+{1\over i}{\partial a_0^{(j)}\over\partial p_j}\cr
}\leqno(b.8)$$
The first equation (b.8) yields $a_0^{(j)}(x,p,E)=b_0(p,E)\lambda_j(x,p,E)$ and the second one can be solved iff
$$\Sum_j\lambda_j(-\partial_p\psi(p),p){\partial b_0\over\partial p_j}(p)
+{\partial \lambda_j\over\partial p_j}(-\partial_p\psi(p),p)b_0(p)=0\leqno(b.9)$$ 
where we have dropped parameter $E$ from the notations. Taking derivative of (b.6) with respect to $x_k$ yields
$$\partial_{x_k}V(x)= \Sum_j\partial_{x_k}\lambda_j(x,p,E)(x_j+\partial_{p_j}\psi(p))+\lambda_k(x,p,E)\leqno(b.10)$$
and evaluating at $x+\partial_p(\psi)=0$, 
$$\lambda_k(-\partial_p\psi, p)=\partial_{x_k}V(-\partial_p\psi)\leqno(b.11)$$
We differentiate again (b.10) with respect to $x_\ell$, and evaluate at $x+\partial_\psi(p)=0$, so we get~:
$$\partial_{x_k}\lambda_\ell(-\partial_p\psi, p)+\partial_{x_\ell}\lambda_k(-\partial_p\psi, p)=
{\partial^2 V\over\partial x_k\partial x_\ell}(-\partial_p\psi)\leqno(b.13)$$
At last, we take derivative of (b.10) with respect to $p_\ell$, and evaluate at $k=\ell$, $x+\partial_p(\psi)=0$,
to obtain~:
$$0=\partial_{p_k}\lambda_k(-\partial_p\psi, p)+\Sum_j\partial_{x_k}\lambda_j(-\partial_p\psi,p)
{\partial^2 \psi\over\partial p_k\partial p_j}(p)$$
so summing over $k$ and using  (b.13) gives
$$0=\partial_{p_k}\lambda_k(-\partial_p\psi, p)+\Sum_{j,k}
{\partial^2 \psi\over\partial p_k\partial p_j}(p){\partial^2 V\over\partial x_k\partial x_j}(-\partial_p\psi)$$
which we insert into (b.9) and using  (b.11) we get again~:
$${1\over 2}\Sum_j{\partial V\over \partial x_j}(-\partial_p\psi){\partial b_0\over \partial p_j}(p)-
\bigl[\Sum_{j<k}
{\partial^2 \psi\over\partial p_k\partial p_j}{\partial^2 V\over\partial x_k\partial x_j}
+{1\over 2}\Sum_k
{\partial^2 \psi\over\partial p_k^2}{\partial^2 V\over\partial x_k^2}\bigr]b_0(p)
=0\leqno(b.14)$$
This can be compared  with (b.2) and again be written in an invariant form as 
$${\cal L}_{X_H|_{\Lambda}}(b_0(p,E)|dp|^{1/2})=0\leqno(b.15)$$
Since ${\partial V\over\partial x_1}(x_0)\neq 0$, this transport equation can be solved with
arbitrary initial condition at $p=p_0$. Again, since  $X_H$ is tangent to $\Lambda$, and  the Lie derivative
invariant under diffeomorphism, this equation holds  iff the pullback of $b_0(p,E)|dp|^{1/2}$ via the projection $\pi_p$
is invariant under the flow of $X_H$. (Here, given a lagrangian immersion $\iota:\Lambda\to T^*M$, identifying locally
$T^*M$ with ${\bf R}^d_{(x,p)}$, we denote by $\pi_p$ the projection on ${\bf R}^d_p$.~)

Once we have determined $b_0(p,E)$, we can solve the first equation (b.8), and so (b.7) holds up to 
first order terms in $h$. Introducing a cutoff in (b.5) around $p_0$, and integrating (b.7) over $p$
in the neighborhood of $p_0$, we find $(H(x,hD_x)-E)u(x,E,h)={\cal O}(h^2)$.

Now we pass to the general situation where rank $d\pi_x(y_0)\leq d-1$ and $d\pi_p(y_0)\leq d-1$. Then 
$\Lambda$ can have a more complicated singularity, e.g. a Whitney crease at $y_0\in {\cal F}_0$ when $d=2$,
or a critical point of the moment map (which we have excluded in our special problem, but occurs when
looking for instance at the critical energy of an harmonic oscillator, see e.g. [KaRo], or more general critical
points [Vu], [CdVVu], \dots.~) From now on we drop for simplicity energy parameter $E$, and remember that everything could
depend on one or several parameters.

Recall from the discussion before Example a.6 that 
$\Lambda$ can be parametrized nar $y_0$ by a non-degenerate phase function $S(p,x)=\langle p,x\rangle+H(p)$.
More generally $\Lambda$ can be parametrized by a Morse function. By this we mean a smooth function $S(x,\theta)$, 
defined near $(x_0,\theta_0)\in{\bf R}^d\times{\bf R}^\ell$ 
(for some $\ell$ which is not necessarily the number $d$ of phase variables, but we can choose $\ell\leq d$)
with $p_0=dS(x_0,\theta_0)$, $d_{(x,\theta)}S(x_0,\theta_0)\neq 0$, and 
$dS'_{\theta_1}\cdots,dS'_{\theta_\ell}$ linearly independant on the critical set
$C_S=\{(x,\theta)\in{\bf R}^d\times{\bf R}^\ell : d_\theta S=0\}$, such that the lagrangian immersions  
$\iota:\Lambda\to T^*M$ and 
$\iota_S:C_S\to T^*M$, $(x,\theta)\mapsto(x,d_xS(x,\theta))$ have the same image 
near $(x_0,p_0)$, which we identify naturally 
with $\Lambda$. We set $\Sigma_S=\{(x,d_xS(x,\theta)): (x,\theta)\in C_S\}\subset\Lambda$.
Such a phase $S(x,\theta)$ satisfies $\det d^2_{(x,\theta)}S\neq 0$, and 
we set $d^2_{(x,\theta)}S(x,\theta)=S''(x,\theta)$ (all derivatives).
For an amplitude with compact support $a(x,\theta,h)$, we define as in (b.5) the lagrangian distribution 
$$I(a,S)(x,h)=(2\pi h)^{-\ell/2}\int e^{iS(x,\theta)/h}a(x,\theta,h)d\theta\leqno(b.17)$$
and consider the $h$-Fourier transformation 
$${\cal F}_hI(a,S)(\xi,h)=(2\pi h)^{-(d+\ell)/2}\int\int e^{i(S(x,\theta)-x\xi)/h}a(x,\theta,h)dx d\theta$$
For $\xi$ near $p_0$ we may expand ${\cal F}_hI(a,S)(\xi,h)$ by stationary phase, since the critical point
$(x(\xi), \theta(\xi))$, with $(x(\xi_0), \theta(\xi_0))=(x_0,\theta_0)$ is non  degenerate. The critical value
equals 
$$\Phi_S(\xi)=S(x(\xi), \theta(\xi))-x(\xi)\xi\leqno(b.18)$$
and we find~:
$${\cal F}_hI(a,S)(\xi,h)=e^{i\Phi_S(\xi)/h}|\det S''|^{-1/2}e^{i\pi(\sgn S'')/4}\bigl(a_0(x(\xi), \theta(\xi))
+{\cal O}(h)\bigr)\leqno(b.19)$$
It is also well known that ${\cal F}_hI(a,S)(\xi,h)$ does only depends on $\Lambda$ near $(x_0,p_0)$ and not on the 
particular choice of $S$ realizing the embedding $\iota:\Lambda\to T^*M$. 
More precisely, if $\widetilde S(x,\widetilde\theta)$ is another phase function as above, such that 
$p_0=d_x\widetilde S(x_0,\widetilde\theta_0)$, there exists an amplitude
$\widetilde a(x,\theta,h)$ such that formally (i.e. microlocally near $(x_0,p_0)$,~)
$I(a,S)(x,h)=I(\widetilde a,\widetilde S)(x,h)$.
The convenient framework to express this invariance is described in term of the frequency set, see  [GuSt], [Iv], \dots
but we can (roughly) understand this by thinking of an equality between asymptotic sums in $h$,
or even at the first order in $h$. 
See e.g. [BaWe,Theorem 4.18] for a proof of equivalence among the Morse family~; 
from this follows actually Darboux-Weinstein Theorem, that we recalled in Sect.1.
We also call $S(x,\theta)$ a non-degenerate phase function in the sense of H\"ormander. 

Choose $\xi=(\xi_1,\cdots, \xi_d)$ as local coordinates on $\Lambda$
and define a half density in the local charts $(C_S,\iota_S)$ near $(x_0,p_0)$ by
$\sqrt{\delta_S}=|\det S''|^{-1/2}|d\xi|^{1/2}$. We notice that if $I(a,S)(x,h)\equiv I(\widetilde a,\widetilde S)(x,h)$
microlocally near $(x_0,p_0)$, then
$$e^{i\Phi_{\widetilde S}(\xi)/h}\widetilde a_0x(\xi), \theta(\xi)\sqrt{\delta_{\widetilde S}}=
e^{i\Phi_S(\xi)/h}a_0e^{i\pi(\sgn S''-\sgn\widetilde S'')/4}\sqrt{\delta_S}\leqno(b.22)$$
Let $\Lambda_S=\iota_S(C_S)\subset\Lambda$  be the image of the chart $(C_S,\iota_S)$. 
The (oscillating) principal symbol of $u$ in $\Lambda_S$ is then defined as 
$$e^{i\Phi_S(\xi)/h}A_0(\xi)=e^{i\Phi_S(\xi)/h}e^{i\pi\sgn S''/4}a_0\bigl(x(\xi),\theta(\xi)\bigr)
\sqrt{\delta_S}\leqno(b.23)$$ 
To determine the principal symbol
$a_0(x(\xi), \theta(\xi))$, we proceed essentially as above, computing $H(x,hD_x)$ $u(x,h)$ by inverse
Fourier transform. So if the lagrangian distribution $u$ has (oscillating) principal symbol $e^{i\Phi_S(\xi)/h}A_0(\xi)$,
then $H(x,hD_x))u(x,h)$, which can be computed as in (b.14),
has (oscillating) principal symbol $e^{i\Phi_S(\xi)/h}{1\over i}{\cal L}_{X_H|_\Lambda}A_0(\xi)$.
(See [DuH\"o] for the case of a classical PDO, and also [Iv] that takes care of specific details relative to the
semi-classical framework.~) 
To have $H(x,hD_x)u(x,h)=0$ mod ${\cal O}(h^2)$, it suffices to solve
$$e^{i\Phi_S(\xi)/h}{1\over i}{\cal L}_{X_H|_\Lambda}A_0(\xi)|d\xi|^{1/2}=0\leqno(b.25)$$ 
and still in the case where $\Lambda$ is an integral manifold of $H$, this equation is satisfied whenever the pullback of the half-density
$A_0(\xi)$ in the chart $(C_S,\iota_S)$ is invariant under the flow of $X_H$. Let us summarize our review in the~:
\medskip
\noindent {\bf Theorem b.1}: Let $\Lambda$ as above be an integral manifold of $H(p,x)$ in energy surface $H(p,x)=E$.
Assume $H(p,x)=p^2+V(x)$ so that 
$H(x,hD_x)=-h^2\Delta+V(x)$, or else $H(x,hD_x,h)$ is the $h$-Weyl quantization of a symbol of the form $H(x,p,h)=H(p,x)+h^2H(p,x)$
(i.e. without subprincipal symbol $H_1(p,x)$.~) Then near any $y_0=(p_0,x_0)\in\Lambda$ there is a local 
chart $(C_S,\iota_S)$ and a half density $\sqrt{\delta_S}$ as above such that the lagrangian distribution $u$ in (b.17)
with (oscillating) principal part $e^{i\Phi_S(\xi)/h}A_0(\xi)$ given by (b.23) solves $(H(x,hD_x,h)-E)u(x,h)={\cal O}(h^2)$. 
\smallskip
In particular, let $y\in\Lambda$ be such that rank $d\pi|_\Lambda(y)=k\leq d-1$ is locally constant, then 
because $\Lambda$ is lagrangian, there is a partition of variables
(possibly after renumerotation of coordinates) $x=(x_1,x_2)\in{\bf R}^k\times{\bf R}^{d-k}$, and
$p=(p_1,p_2)\in{\bf R}^k\times{\bf R}^{d-k}$, such that rank $d\widetilde \pi=d$, where $\widetilde \pi:\Lambda\to{\bf R}^d$,
$(x,p)\mapsto(x_1,p_2)$. Then it is well known 
that $u(x,h)$ in (b.17) can be rewritten in the same form with $\theta$ replaced by $p_2$ and 
$S\circ\iota^{-1}\circ\widetilde\pi(x_1,p_2)=T(x_1,p_2)+\langle x_2,p_2\rangle$. 
In particular, near $y$,
$$\Lambda=\Lambda_T=\{(x,p): x_2=-{\partial T\over\partial p_2}(x_1,p_2),
p_1={\partial T\over\partial x_1}(x_1,p_2)\}\leqno(b.27)$$
and the number $\ell$ of phase variables can be reduced to $d-k$.  We interprete $T(x_1,p_2)$ as a
partial Legendre transformation of $\phi(x)$ or $\psi(p)$. See [Iv,Prop.1.2.5], [DoZh], \dots 
If we content computing Maslov indices, one could restrain to such phase functions as we shall explain below. 

Now we want to extend the previous procedure to more general $h$-PDO's,  
in particular for (Weyl quantization of) $H(x,hD_x,h)=H_0(x,hD_x)+hH_1(x,hD_x)+\cdots$, 
in this case $\sigma_H(x,p)=H_1(x,p)$ is called the sub-principal symbol of $H$.
Of course, we need not assume the particular form $-h^2\Delta+V(x)$ for $H_0$. When
$H(x,hD_x,h)$ is the Weyl quantization of symbol $H(x,p)=H_0(x,p)+hH_1(x,p)+\cdots$, then 
$\sigma_H(x,p)=H_1(x,p)$, and there are formulas relating subprincipal symbols for different quantizations, see e.g. [Iv]. 
Note that the sub-principal symbol, together with the principal symbol, are invariantly defined.

Now if $u$ is a lagrangian distribution as above, 
then in the chart $(C_S,\iota_S)$, the distribution $H(x,hD_x,h)u(x,h)$
has (oscillating) principal symbol 
$$e^{i\Phi_S(\xi)/h}\bigl({1\over i}{\cal L}_{X_{H_0}|_\Lambda}+\sigma_H(x(\xi),p(\xi)\bigr)A_0(\xi)$$
and again equation
$$\bigl({1\over i}{\cal
L}_{X_{H_0}\big|_\Lambda}+\sigma_H(x(\xi),p(\xi))\bigr)A_0(\xi)|d\xi|^{1/2}=0\leqno(b.29)$$
can be solved along the integral curves of $X_{H_0}|_\Lambda$. 
Where $\Lambda$ is projectable on $M$, we can take $x$ as local coordinates on $\Lambda$, and this equation takes the form 
$${1\over 2i}\Sum_j{\partial H_0\over\partial p_j}|_\Lambda{\partial\over\partial x_j}A_0^2(x)+\bigl(\sigma_H|_\Lambda\bigr) A_0^2(x)=0$$
but near a simple fold, it can be computed as in (b.14), and more generally using the procedure leading to Theorem b.1.
This gives on $\Lambda$~:
$$A_0(\xi)=\Const |\det {\partial \xi\over\partial\varphi}|^{-1/2}
\exp \bigl[-i\int^t\sigma_H(p(\xi)x(\xi))ds\bigr]\leqno(b.30)$$
Assume now that $\Lambda$ is a torus with quasi-periodic flow of frequencies $\omega$ satisfying some
diophantine condition we give (b.30) a simpler form. Namely expand 
$\sigma_H(p(\xi),x(\xi))\circ(\omega s+\varphi)$ as a Fourier series, and integrate term by term  
which gives $\int^t\sigma_H(p(\xi),x(\xi))\circ(\omega s+\varphi)ds=\langle \sigma_H\rangle t + G(\varphi)$, 
where $G$ is a smooth periodic function on the torus, satisfying $\langle G\rangle=0$,
and $\langle \sigma_H\rangle$, again, is the spatial average of $\sigma_H$  over $\Lambda$. Of course, 
this is also the time average over the almost periodic orbit. In (b.30) we can factor out the periodic part,
which is the same as the solution of (b.25), multiplied by $e^{-iG(\varphi)}$. This will give the invariant density
(since its Liouville class is zero.~) The other factor, $e^{-i\langle \sigma_H\rangle t }$ is multivalued.
So far we rewrite (b.30) as 
$$A_0(\xi(\omega t+\varphi))=\Const |\det {\partial \xi\over\partial\varphi}|^{-1/2}
\exp [-iG(\omega t+\varphi)]\exp [-i\langle \sigma_H\rangle t]\leqno(b.31)$$
This expression is computed along the integral curves of $X_{H_0}$~; of course it assumes a particularly simple form 
when $\sigma_H\rangle=0$. 
\medskip
\noindent {\bf C. Maslov quantization condition for a lagragian immersion of the torus}.
\medskip
At last we investigate the global properties of the lagrangian distribution $u(x,h)$ 
we have so far constructed locally. Choose $\xi=(\xi_1,\cdots, \xi_d)$ as local coordinates on $\Lambda$
and define a half density in the local charts $(C_S,\iota_S)$ near $(p_0,x_0)$ by
$\delta_S=|\det S''|^{-1}|d\xi|$. We notice that if $I(a,S)(x,h)=I(\widetilde a,\widetilde S)(x,h)$
microlocally near $(p_0,x_0)$, then
$$e^{i\Phi_{\widetilde S}(\xi)/h}\widetilde a_0\sqrt{\delta_{\widetilde S}}=
e^{i\Phi_S(\xi)/h}a_0\sqrt{\delta_S}e^{i\pi(\sgn S''-\sgn\widetilde S'')/4}\leqno(c.1)$$
Let $\Lambda_S=\iota_S(C_S)$  be the image of the chart $(C_S,\iota_S)$. 
We denote by $\Omega_{1/2}$ the bundle of half-densities on $\Lambda$ with  
transition functions 
$\exp i\pi\bigl( \sgn S''_{(x,\theta),(x,\theta)}-\sgn S''_{(x',\theta),(x',\theta)}\bigr)/4$ 
for change of coordinates $x\mapsto x'$ in $\Lambda_S$. We denote also by ${\bf L}$ 
Maslov bundle on $\Lambda$ with transition functions $\exp i\pi(\sgn S''-\sgn\widetilde S'')/4$
for changes of function $S$ in $\Lambda_S\cap\Lambda_{\widetilde S}$.
With $I(a,S)$ we associate the section of $\Omega_{1/2}\otimes{\bf L}$ of the form 
$$e^{i\Phi_S(\xi)/h}A_0(\xi)=e^{i\Phi_S(\xi)/h}e^{i\pi\sgn S''/4}a_0\bigl(x(\xi),\theta(\xi)\bigr)
\sqrt{\delta_S}\leqno(c.2)$$ 
by a partition of unity subordinated to the covering of $\Lambda$ by the local charts $\Lambda_S$.

Maslov canonical operator is defined as the inverse Fourier transform which to
any section $e^{i\Phi_S(\xi)/h}$ $A(\xi, h)$ of $\Omega_{1/2}\otimes{\bf L}$ as in (c.2),
assigns a lagrangian distribution $u(x,h)$. We call $e^{i\Phi_S(\xi)/h}A(\xi, h)$ the (oscillating) 
principal symbol of $u$.
The interesting case is when $\Lambda$ is a compact manifold. Namely, if the open sets $\Lambda_S$ cover $\Lambda$,
then the functions 
$\Phi_{S,\widetilde S}(\xi)=\Phi_S(\xi)-\Phi_{\widetilde S}(\xi)=S(x_0,\theta_0)-\widetilde S(x_0,\theta'_0)$, for
$\iota_S(x_0,\theta_0)=\iota_{\widetilde S}(x_0,\theta'_0)=(x_0,p_0)\in\Lambda_S\cap\Lambda_{\widetilde S}$, 
are necessarily constant on $\Lambda_S\cap\Lambda_{\widetilde S}$, defining an element $\Phi\in H^1(\Lambda;{\bf R})$.
Actually, by De Rham isomorphism, $\Phi\approx\iota^*(pdx)$. In the same way, if the covering $\Lambda_S$
is sufficiently smooth, the transition functions $\sigma_{S,\widetilde S}=e^{i\pi(\sgn S''-\sgn\widetilde S'')/4}$ 
are constant on $\Lambda_S\cap\Lambda_{\widetilde S}$, defining an element $\alpha\in H^1(\Lambda;{\bf Z})$, 
and $\alpha=\mu(L,L_2)$ is Maslov class of $(\Lambda,\iota)$.
Of course, these transition functions are nothing but those we computed in Part B. 
Maslov quantization condition expresses the fact that 
$$(\Phi_S(\xi)-\Phi_{\widetilde S}(\xi))/h\equiv \pi(\sgn S''-\sgn\widetilde S'')/4 \mod 2\pi \ \hbox{on} \ 
\Lambda_S\cap\Lambda_{\widetilde S}$$
which we write as a condition on $\Omega_{1/2}\otimes{\bf L}$~:
$$\Phi/(2\pi h)+ \alpha/4=0\mod H^1(\Lambda;{\bf Z}) \leqno(c.3)$$
The values of $h$ satisfying (c.3) are generally of the form ${1\over h}=an+b, n\in{\bf Z}$, and Maslov index
$\alpha=(2,\cdots,2)$ for a torus. Global topological properties of 2-d tori are considered in [O-de-AlHa]. In particular,
we can have $\alpha=(2,0)$ in 2-d.

At last, we point out how to correct Maslov quantization condition when $H$ has a 
sub-principal symbol, and the frequencies of the torus are diophantine. In (b.31) $e^{-i\langle \sigma_H\rangle t }$ is multivalued, and
contributes to Maslov condition by a constant term. So in case of a sub-principal symbol, we need to modify (c.3) as~:
$$\Phi/(2\pi h)+\alpha/4 +\langle \sigma_H\rangle(1,\cdots,1)=0\mod H^1(\Lambda;{\bf Z}) \leqno(c.4)$$
This situation is of course well-known, and related to the sub-principal form for an integrable system
(see [Vu] and references therein.~)

\bigskip
\noindent {\bf References}
\medskip

\noindent [Ar]  V.Arnold {\bf 1}. 
On a characteristic class entering into conditions of quantization. Funct. Anal. Appl. 1, p.1-13, 1967.
{\bf 2}. M\'ethodes math\'ematiques de la M\'ecanique classique. Editions Mir, Moscou, 1976. 
{\bf 3}. Geometrical methods in the theory of ordinary differential equations. Springer, 1983.

\noindent [Au] M. Audin. Les syst\`emes hamiltoniens et leur int\'egrabilit\'e. Soc. Math de France, Cours Sp\'ecialis\'es (8), 2001. 

\noindent [BNe] I.Babenko, N.Nekhoroshev. On complex structures on 2-d tori admitting metrics with nontrivial quadratic integral.
Math. Notes, 58(5) p.1129-1135, 1995.

\noindent [BaWe] S.Bates, A.Weinstein. Lectures on the geometry of quantization. Berkeley Math. Lect. Notes 88,
American Math. Soc. 1997.

\noindent [BeDoMa] V.Belov, S.Dobrokhotov, V.Maksimov. Explicit formulas for  generalized action-angle variables
in a neighborhood  of an isotropic torus and their applications. Theor. Math. Phys., 135(3),p.765-791, 2003.

\noindent [Ch] S.Chandrasekhar. The mathematical theory of black holes.  
International Series of Monographs on Physics, 69. The Clarendon Press, Oxford University Press, New York,  
1983.

\noindent [CdV]  Y.Colin de Verdi\`ere,  {\bf 1}. Modes et quasi-modes sur les vari\'et\'es riemanniennes. Inventiones Math.
43, p.15-52, 1977. {\bf 2}. M\'ethode de moyennisation en M\'ecanique
semi-classsique. Journ\'ees Equations aux D\'eriv\'ees partielles, Expos\'e No 5, Saint Jean de Monts, 1996.

\noindent [CdVVu] Y.Colin de Verdi\`ere, S.Vu Ngoc. Singular Bohr-Sommerfeld rules for 2-d integrable systems.
Annales Scient. Ecole Normale Sup\'erieure, 2003.

\noindent [deG] M.de Gosson. Symplectic Geometry and Quantum Mechanics. Operator Theory Advances and Applications, Birkh\"auser, 2006.

\noindent [DoZh] S.Dobrokhotov, A.Shafarevich. ``Momentum'' tunneling between tori and the splitting of 
eigenvalues of the Laplace-Beltrami operator on Liouville surfaces. Math. Phys. Anal. Geometry 2,
p.141-177, 1999. 

\noindent [DoZh] S.Dobrokhotov, P.Zhevandrov. Asymptotic expansions and the Maslov canonical operator in the 
linear theory of water waves I. Russian J. Math. Phys. Vol.10 (1), p.1-31, 2003.

\noindent [DuFoNo] B. Dubrovin, A.Fomenko, S.Novikov. Modern Geometry-Methods and applications. Vol. I, II, III.
Springer, 1985. 

\noindent [Dui] J.J. Duistermaat. {\bf 1}. Oscillatory integrals, Lagrange immersions and unfolding of singularities. 
Comm. Pure Appl. Math. 27, p.207-281, 1974. {\bf 2}. Fourier Integral Operators, Birkh\"auser, Boston, 1995.
{\bf 3}. On the Morse index in variational calculus. Adv. Math. 21, p.173-195, 1976. 

\noindent [EasMat] M.Eastwood, V.M.Matveev. Metric connexions in projective geometry. Preprint.

\noindent [FedMas] M.V.Fedoriuk, V.P.Maslov. Semi-classical approximation in Quantum Mechanics. D.Reidel, 1981.
 
\noindent [Fo] A.Fomenko. Integrability and non-integrability in Geometry and Mechanics. Kluwer Acad. Publ. 1988.

\noindent [GuSt] V.Guillemin, S.Sternberg. Geometric asymptotics. American Math. Soc. Surveys, 14, Providence, Rhode Island, 1977.

\noindent [H\"a] D.H\"afner. Creation of fermions by rotating black holes. Preprint arXiv:math.AP/061 2501v1

\noindent [HeSj] B. Helffer, J. Sj\"ostrand. R\'esonances en limite semi-classique. Bull. Soc. Math. France
114(3), M\'emoire No 24/25, 1986. 

\noindent [HieGrDoRa] J.Hietarinta, B.Grammaticos, D.Dorizzi, A.Ramani. Phys. Rev. Lett. V.53, 1707, 1984.

\noindent [HiSjVu] M.Hitrik, J.Sj\"ostrand, S.Vu-Ngoc. Diophantine tori and spectral asymptotics for non-self adjoint 
operators.

\noindent [H\"o] L.H\"ormander {\bf 1}. The Analysis of Linear Partial Differential Operators, III. Springer-Verlag, 
Berlin, 1985. {\bf 2}. Fourier Integral Operators I. Acta Math. 127, p.79-183, 1971.

\noindent [Iv] V.Ivrii.  Microlocal analysis and precise spectral asymptotics. Springer-Verlag, Berlin,  1998.

\noindent [KaRo] N.Kaidi, M.Rouleux. Quasi-invariant tori and semi-excited states for Schr\"odinger operators I.
Asymptotics. Comm. Part. Diff. Eq. 27, p.1695-1750, 2002.

\noindent [KMS] D.V.Kosygin, A.A.Minasov, Ya.G.Sinai,  
Statistical properties of the spectra of the Laplace-Beltrami operators on Liouville surfaces,
Russ. Math. Surv. Vol.~48, no.~4, p.~1-142, 1993.

\noindent [Laz] V.Lazutkin. KAM theory and semi-classical approximation to eigenfunctions. Springer, 1993.

\noindent [Le] J.Leray. Analyse lagrangiennne et M\'ecanique Quantique. S\'eminaire EDP Coll\`ege de France, 1976-77.

\noindent [Li] R.G. Littlejohn. Semiclassical trace formulas. J. Math. Phys. 31(12), p.2952-2977, 1990.

\noindent [Ma] V.P.Maslov. {\bf 1}. Th\'eorie des perturbations et m\'ethodes asymptotiques. Dunod, Paris, 1972.
{\bf 2}. Operational Methods. Moscow: Mir~Publ. 1973.

\noindent [Mat] V.S.Matveev. The asymptotic eigenfunctions of the operator $\nabla D(x,y)\nabla$ corresponding to
Liouville metrics and waves on water captured by bottom irregularities. Math. Notes, 64(3), p.357-363, 1998.

\noindent [MatTo] V.S.Matveev, A.Topalov. {\bf 1}. Quantum integrability... Math. Zeitschrift, 238(4), p.833-860, 2001.
{\bf 2}. Geometric equivalence... Geom. Dedicata, 96, p.91-115, 2003.

\noindent [MeSj] A.Melin, J.Sj\"ostrand. Fourier integral operators with complex valued phase functions. Springer
Lect. Notes in Math. 459, p.120-223, 1974. 

\noindent [Mi] H.Mineur. R\'eduction des syst\`emes m\'ecaniques \dots
J. Math. Pures Appl., 15, p.385-389, 1936.

\noindent [O-de-AlHa] A.M. Ozorio de Almeida, J. H. Hannay. Geometry of 2-d
tori in phase space~: projections, sections and the Wigner function. Ann. Phys.
138, p.115-154, 1982

\noindent [Po] G.Popov. Invariant tori, effective stability, and quasi-modes with exponentially small error terms. {\bf 1}.
Birkhoff normal forms, Ann. Henri Poincar\'e 1, p.223-248, 2000.  {\bf 2}.
Quantum Birkhoff normal forms, Ann. Henri Poincar\'e 1, p.249-279, 2000.  

\noindent [Se] E.Selivanova. Orbital isomorphisms of Liouville systems on a 2-d torus. Mat. Sbornik 186(10), p.141-160,
1995. 

\noindent [So] J.M.Souriau, Construction explicite de l'indice de Maslov. Applications, {\it in}:
Group theoretical methods in Physics, Nijmegen. Lect. Notes in Physics 50, Springer, 1976.
 
\noindent [Ts] A.Tsiganov. The Maupertuis principle and canonical transformations of the extended phase space.
J. Nonlinear Math. Phys. 8(1), p.157-182, 2001.

\noindent [Vu] S.Vu Ngoc. Bohr-Sommerfeld conditions for integrable systems with critical manifolds of focus-focus type.
Comm. Pure Appl. Math.,53(2), p.143-217, 2000.

\noindent [We] A.Weinstein. {\bf 1}. On Maslov quantization condition, {\it in}:  Fourier Integral Operators and 
Partial Differential Equations. J.Chazarain, ed. Lecture Notes in Math. 469, Springer, p.361-372, 1974.  
{\bf 2}.  Symplectic manifolds and their lagrangian submanifolds, Adv. Math. 6, p.329-346, 1971.

\noindent [Ze] S.Zelditch. The inverse spectral problem for surfaces of revolution. J. Differential Geom. 49, p.207-264, 1998.

\bye